\documentclass[aps,prd,superscriptaddress,showpacs,nofootinbib,10pt,notitlepage]{revtex4-2}

\usepackage{graphicx}
\usepackage{amsmath,amssymb}
\usepackage{bbm}
\usepackage[dvipsnames]{xcolor}
\usepackage[utf8]{inputenc}
\usepackage[colorlinks=true,allcolors=BlueViolet,hyperfootnotes=false]{hyperref}
\usepackage{lmodern}
\usepackage[T1]{fontenc}
\usepackage[inline]{enumitem}
\usepackage{accents}
\usepackage{slashed}
\usepackage{orcidlink}
\usepackage{natbib}
\usepackage{stackengine}

\usepackage[normalem]{ulem}

\usepackage{tikz}
\usetikzlibrary{shapes.geometric}

\graphicspath{{Images/}}

\newcommand{\be}{\begin{equation}} \newcommand{\ee}{\end{equation}}
\newcommand{\ba}{\begin{array}{c}} \newcommand{\ea}{\end{array}}
\newcommand{\bea}{\begin{eqnarray}} \newcommand{\eea}{\end{eqnarray}}

\newcommand{\ols}[1]{\mskip.5\thinmuskip\overline{\mskip-.5\thinmuskip {#1} \mskip-.5\thinmuskip}\mskip.5\thinmuskip} 
\newcommand{\olsi}[1]{\,\overline{\!{#1}}} 

\def\XXint#1#2#3{{\setbox0=\hbox{$#1{#2#3}{\int}$}
     \vcenter{\hbox{$#2#3$}}\kern-.5\wd0}}

\begin{document}

\title{\boldmath Charge-conjugation asymmetry and molecular content: the \texorpdfstring{$D_{s0}^\ast(2317)^\pm$}{Ds(2317)+} in matter }

\newcommand{\ific}{Instituto de F\'{\i}sica Corpuscular (centro mixto CSIC-UV),
Institutos de Investigaci\'on de Paterna,
C/Catedr\'atico Jos\'e Beltr\'an 2, E-46980 Paterna, Valencia, Spain}
\newcommand{\ice}{Institute of Space Sciences (ICE, CSIC), Campus UAB,  Carrer de Can Magrans, 08193 Barcelona, Spain}
\newcommand{\ieec}{Institut d'Estudis Espacials de Catalunya (IEEC), 08860 Castelldefels (Barcelona), Spain}
\newcommand{\fias}{Frankfurt Institute for Advanced Studies, Ruth-Moufang-Str. 1, 60438 Frankfurt am Main, Germany}

\author{V.~Montesinos\orcidlink{0000-0002-6186-2777}}
\email{Victor.Montesinos@ific.uv.es}
\author{M.~Albaladejo\orcidlink{0000-0001-7340-9235}}
\email{Miguel.Albaladejo@ific.uv.es}
\author{J.~Nieves\orcidlink{0000-0002-2518-4606}}
\email{jmnieves@ific.uv.es}
\affiliation{\ific}
\author{L.~Tolos\orcidlink{0000-0003-2304-7496}}
\email{tolos@ice.csic.es}
\affiliation{\ice}
\affiliation{\ieec}
\affiliation{\fias}

\definecolor{citecolor}{rgb}{0.15,0.15,0.60}

\begin{abstract}
We analyze the modifications that a dense nuclear medium induces in the $D_{s0}^\ast(2317)^\pm$ and $D_{s1}(2460)^\pm$. In the vacuum, we consider them as isoscalar $D^{(*)}K$ and $\olsi{D}{}^{(*)}\olsi{K}$ $S$-wave bound states, which are dynamically generated from  effective interactions that lead to  different Weinberg compositeness scenarios. Matter effects are incorporated through the two-meson loop functions,  taking into account the self energies that the $D^{(*)}$, $\olsi{D}{}^{(*)}$, $K$, and $\olsi{K}$ develop when embedded in a nuclear medium. Although particle-antiparticle [$D^{(\ast)}_{s0,s1}(2317,2460)^+$ \textit{versus} $D^{(\ast)}_{s0,s1}(2317,2460)^-$] lineshapes are the same in the vacuum, we find extremely different density patterns in matter. This charge-conjugation asymmetry mainly stems from the very different kaon and antikaon interaction with the nucleons of the dense medium. We show that the in-medium lineshapes found for these resonances strongly depend on their $D^{(*)}K$/$\olsi{D}{}^{(*)}\olsi{K}$ molecular content, and discuss how this  novel feature can be used to better determine/constrain the inner structure of these exotic states.
\end{abstract}

\maketitle

\section{Introduction} \label{sec:intro}

The $D_{s0}^\ast(2317)^\pm$ state was first reported by the BaBar Collaboration in 2003 \cite{BaBar:2003oey}, and was a little after confirmed by CLEO in a Ref.~\cite{CLEO:2003ggt} where the observation of the $D_{s1}(2460)^\pm$ was also claimed. These resonances lie far below the predictions for the two expected broad $P$-wave $c\olsi s$ mesons~\cite{Godfrey:1985xj, Godfrey:1986wj,Zeng:1994vj,Gupta:1994mw,Ebert:1997nk,Lahde:1999ih, DiPierro:2001dwf}, while they are located near the $D K$ and $D^\ast K$ thresholds, respectively, at about $45\,\text{MeV}$ below them. Both states are isoscalars  [$I(J^P)=0(0^+)$ and $I(J^P)=0(1^+)$, respectively]  and thus strong isospin-violating decays are possible only to 
the isovector channels $D_s^{(*)}\pi $ leading to very small widths ($\lesssim 4\,\text{MeV}$ at 95\% confidence level \cite{ParticleDataGroup:2022pth}). 

The $D_{s0}^\ast(2317)$ and $D_{s1}(2460)$ states were, together with the $\chi_{c1}(3872)$, some of the first ever exotic hadronic states discovered. More specifically, the $D_{s0}^\ast(2317)$ and $D_{s1}(2460)$ are exotic in the sense that they give rise to three puzzles~\cite{Du:2017zvv}, 
\begin{enumerate}
    \item The masses for both states are significantly lower than the predictions stemming from the Godfrey and Isgur quark model~\cite{Godfrey:1985xj}, which was incredibly successful at the time (and even now).
    \item The splitting  between the $D_{s1}$ and the $D_{s0}^\ast$ is equal (up to a few MeV) to the mass difference between the $D^*$ and $D$ mesons. 
    \item The mass of the $D_{0}^\ast(2300)$ state, which does not contain any strange quark, is found to be larger than that of the $D_{s0}^\ast(2317)$,   even though one should expect $c \olsi s$ states to be in general heavier than  $c\olsi \ell$ ($\ell=u,d$) ones (hierarchy puzzle).
\end{enumerate}
These problems are naturally solved within the chiral molecular picture, with a double pole structure for the  broad $D_{0}^\ast(2300)$ resonance, and large (dominant)  $D K$ and $D^*K$  components \cite{Weinberg:1965zz,Albaladejo:2022sux} in the  $D^\ast_{s0}(2317)^+$ and $D_{s1}(2460)^+$ cases, respectively. In this scheme \cite{Albaladejo:2016lbb,Du:2017zvv}, the SU(3) $D^{(*)}_{(s)}\phi$ (with $\phi$ a Goldstone boson) $S$-wave  scattering, in the  $J^P=0^+$ and  $1^+$ sectors,  is studied using next-to-leading (NLO)  heavy meson chiral perturbation theory\footnote{An effective Lagrangian that describes the low momentum interactions of mesons
containing a heavy quark with the pseudo-Goldstone bosons $\pi,K$ and $\eta$. It is invariant under both heavy quark spin  and chiral SU(3)$_L$ $\times$ SU(3)$_R$ symmetries~\cite{Wise:1992hn,manohar_wise_2000}.} (HMChPT) unitarized amplitudes, as initially proposed in Ref.\,\cite{Guo:2008gp} with sub-leading low energy constants (LECs) determined in Ref.\,\cite{Liu:2012zya}. The dynamical origin of the  $D_{0}^\ast(2300)$ two-state structure comes  from the light-flavor SU(3) structure of the interaction, and it was found that the lower pole would be the SU(3) partner of the $D^\ast_{s0}(2317)$~\cite{Albaladejo:2016lbb}. 

Nevertheless, we should mention that there exist many works discussing  different scenarios  for the structure of the $D_{s0}^*(2317)$. Thus, conventional $c  \olsi q$ models \cite{Colangelo:2003vg, Dai:2003yg, Narison:2003td, Bardeen:2003kt, Lee:2004gt, Wang:2006bs, Lakhina:2006fy}, tetraquark $cq\olsi q \olsi q$ interactions \cite{Cheng:2003kg, Terasaki:2003qa, Chen:2004dy, Maiani:2004vq, Bracco:2005kt, Wang:2006uba}, molecular heavy-light meson-meson approaches \cite{Barnes:2003dj, Szczepaniak:2003vy, Kolomeitsev:2003ac, Hofmann:2003je, Guo:2006fu, Gamermann:2006nm, Faessler:2007gv, Flynn:2007ki, Guo:2008gp, Guo:2009ct, Liu:2012zya, Guo:2015dha, Albaladejo:2016lbb, Albaladejo:2016hae, Guo:2017jvc} or combinations of conventional quark models plus pure tetraquark or meson-meson molecules~\cite{Browder:2003fk, vanBeveren:2003kd, Ortega:2016mms, Albaladejo:2018mhb} have been suggested. A great effort has also been devoted to studying these states in lattice QCD. Initially Refs.\,\cite{Bali:2003jv,Dougall:2003hv}, which used only interpolators of the type $c \olsi{s}$, reported masses for the $D_{s0}^\ast$ greater than its physical value. Later, the simulations of Refs.\,\cite{Mohler:2012na,Mohler:2013rwa,Lang:2014yfa,Bali:2017pdv} obtained masses consistent with those of the physical $D_0^\ast(2300)$ and $D_{s0}^\ast(2317)$ states by including also two-meson (four-quark) operators. Recently, more complete studies for the $D_{s0}^\ast$ have been performed by the HadSpec Collaboration \cite{Cheung:2020mql} leading to a fair description of the isoscalar $DK$  and isoscalar and isovector $D\olsi{K}$ scattering as well as the $D_{s0}^*(2317)$ from LQCD. Actually, these latter results are in good agreement with those predicted in the unitarized HMChPT model presented in Ref.\,\cite{Albaladejo:2016lbb} (see also Ref.\,\cite{Albaladejo:2018mhb}). The scheme of Ref.\,\cite{Albaladejo:2016lbb} led also to a remarkably good description of the LQCD low-lying levels reported by HadSpec in Ref.\,\cite{Moir:2016srx} in the  $J^P=0^+$  strangeness-isospin $(S,I)=(0,1/2)$, which gave strong support for the existence of two poles in the $D_0^*(2300)$ energy region. This picture is also confirmed by the latest HadSpec results released in  Ref.\,\cite{Gayer:2021xzv} and  by the high-quality data on the $B^- \to
D^+\pi^-\pi^-$ and $B_s^0 \rightarrow \olsi{D}{}^0 K^-\pi^+$ final states provided by the LHCb experiment in Refs.~\cite{LHCb:2016lxy} and
\cite{LHCb:2014ioa}, respectively, and analyzed in Ref.\,\cite{Du:2017zvv}. Given the interest in these resonances, other methods to explore their nature have been proposed, such as the calculation of the femtoscopic correlation functions for the involved channels \cite{Liu:2023uly,Albaladejo:2023pzq,Ikeno:2023ojl,Torres-Rincon:2023qll}.

In summary, the topic of the internal structure of the $D_{s0}^\ast$ has been of interest to the hadronic community for two decades, as it can have a profound impact on the pillars in which the theoretical description of the hadronic spectrum is based. In this work, we aim to study how the properties of the $D_{s0}^\ast(2317)^\pm$ and the $D_{s1}(2460)^\pm$ change when they are embedded in matter, for different Weinberg compositeness scenarios~\cite{Weinberg:1965zz}. The idea is that their $D^{(*)}K$ and $\olsi D{}^{(*)}\olsi K$  molecular components would get renormalized in a different way because the presence of the nuclear medium produces a charge-conjugation asymmetry. Owing to the different nature of the $D^{(*)}N$ and $\olsi D{} ^{(*)}N$ and of the  $KN$ and $\olsi{K}N$ interactions, we expect characteristic changes of the in-medium properties of the $D_{s0}^{\ast}(2317)^+$ and $D_{s1}(2460)^+$ \textit{versus} those of the $D_{s0}^{\ast}(2317)^-$ and $D_{s1}(2460)^-$, which should become increasingly visible as the density increases. The future experimental confirmation of the found distinctive density pattern would give support and help to constrain  the importance of molecular components in these exotic states.\footnote{This is because it is reasonable to think that this density-dependent particle-antiparticle asymmetry would be different for compact  ($c\olsi{q}$ or $cq\olsi q \olsi q$) structures.} Note that a preliminary study of the $D_{s0}^\ast(2317)^+$, but not of the $D_{s0}^\ast(2317)^-$, in dense matter has been performed in Ref.~\cite{Molina:2009zeg}. Also, the impact of a thermal medium on the $D_{s0}^\ast(2317)^\pm$ has been studied in Ref.~\cite{Montana:2020lfi}, starting from the NLO HMChPT scheme described above, paying attention to the evolution of its mass and decay width as functions of temperature, and  to the possibility of chiral-symmetry restoration in the heavy-flavor sector below the transition temperature. However, such a study does not show any asymmetry between $D_{s0}^\ast(2317)^+$ and $D_{s0}^\ast(2317)^-$  because  in a hot pion bath, the properties of both resonances will be identical since $D^{(*)}\pi$ and $\olsi D{}^{(*)}\pi$ as well as $K\pi$ and $\olsi{K}\pi$ interactions are equal in the SU(2) limit.

The present study is similar to our previous analysis of Ref.~\cite{Montesinos:2023qbx}, where we also showed that  the asymmetrical density pattern of the properties of the $T_{cc}(3875)^+$ and $T_{\olsi{c}\olsi{c}}(3875)^-$ inside a nuclear environment could become an interesting tool to disentangle the structure ($ccqq$ compact or $DD^*$ molecular) of the exotic $T_{cc}(3875)^+$ tetraquark.  In addition, we expect larger effects in this work because of the substantial difference between antikaon-nucleon and kaon-nucleon interactions. While the $S$-wave $KN$ interaction is very weak, since the kaon contains an antiquark $\olsi{s}$, and it does not produce any resonance at low  energies, the $\olsi K N$ interaction is quite strong, and the $\Lambda(1405)$ and $\Lambda(1670)$ states can be excited.\footnote{This fact has been used to derive new Bell's inequalities for entangled $K^0\olsi{K}{}^0$ pairs produced in $\phi-$meson decays considering the distinct $K^0 N$ and $\olsi{K}{}^0 N$ interactions~\cite{Bramon:2001tb}. The idea is that if a dense piece of nuclear matter is inserted along the neutral kaon trajectory, by detecting the products from strangeness-conserving strong reactions the incoming state is projected either into $K^0$ ($K^0p \to K^+n$) or into $\olsi{K}{}^0$ ($\olsi{K}{}^0p \to \Lambda \pi^+$, $\olsi{K}{}^0n \to \Lambda \pi^0$, $\olsi{K}{}^0 n \to p K^-$). Given the different magnitude of the corresponding cross sections, the slab of nuclear matter acts as a $K^0$ regenerator, since the probability of disappearance of the neutral antikaon $\olsi{K}{}^0$ is significantly larger.} In particular, the $\Lambda(1405)$  has been the object of study of many works in the literature (see for example Refs.~\cite{Kaiser:1995eg,Oset:1997it,Oller:2000fj,Lutz:2001yb,Garcia-Recio:2002yxy,Jido:2003cb,Garcia-Recio:2003ejq,Oller:2005ig,Hyodo:2007jq,Hyodo:2011ur,Ikeda:2012au,Roca:2013av,Kamiya:2016jqc,Nieves:2024dcz}) and is found to be a very broad quasi-bound state that is well described by a two-pole structure in the scalar-isoscalar ($\pi\Sigma$, $\olsi K N$, $\eta \Lambda$, $K\Xi$) chiral coupled-channel $S$-wave scattering amplitude.

Given the quasi-bound nature of the $\Lambda(1405)$, the possible existence of deeply bound nuclear $K^-$ states was suggested in the work of Akaishi and Yamazaki (AY)~\cite{Akaishi:2002bg}. There, the nuclear ground states of a $K^-$ in $^3$He, $^4$He, and $^8$Be are predicted to be discrete states with binding energies of 108, 86, and 113 MeV and widths of 20, 34, and 38 MeV, respectively. These results are in apparent disagreement with what can be inferred from the previous predictions of Ref.~\cite{Baca:2000ic}. The latter paper employed the self-energy of the $K^-$ meson in nuclear matter calculated in a self-consistent microscopic approach~\cite{Ramos:1999ku}, using the unitarized  $\olsi{K}N$ $T$-matrix in the free space obtained in the LO chiral approach of Ref.~\cite{Oset:1997it}. However,  FINUDA spectrometer claimed  in 2005 \cite{FINUDA:2005lqd} the observation of a kaon-bound state $K^-pp$ through its two-body decay into  $\Lambda$ hyperon and a proton. The binding energy and the decay width were determined to be  $115\pm 9\,\text{MeV}$ and $67\pm 16\,\text{MeV}$, respectively, in line with the expectations of Ref.~\cite{Akaishi:2002bg}. However, some doubts soon arose on the interpretation given by FINUDA~\cite{Oset:2005sn, Magas:2006fn, Weise:2010xn}, and it is theoretically  accepted  that the $\olsi{K}-$nucleus  interaction has little in common with the strongly attractive, local and energy-independent AY  potential of Ref.~\cite{Akaishi:2002bg} used in their calculations of deeply bound $\olsi{K}$ nuclear clusters. Although the effective interaction deduced from chiral SU(3) dynamics is quite attractive, it turns out to be far less attractive than the AY potential in the deep $\olsi{K}N$ sub-threshold region~\cite{Ramos:1999ku, Weise:2010xn} (see also Ref.~\cite{Tolos:2020aln} for a review).

In this work we start by considering  the isoscalar $S$-wave $D^{(\ast)}K$ scattering amplitude and its charge conjugated channel in the vacuum, where the $D_{s0}^\ast(2317)^\pm$ and $D_{s1}(2460)^\pm$ will be generated for different molecular probabilities scenarios, while paying special attention to the comparison with the HMChPT scheme. Afterwards, we will include nuclear medium effects through the heavy-light meson-Goldstone boson loop function, which gets renormalized by the density-dependent $K$, $\olsi K$, $D^{(\ast)}$ and $\olsi D{}^\ast$ spectral functions, with the latter taken from previous works~\cite{Tolos:2009nn,Garcia-Recio:2010fiq,Garcia-Recio:2011jcj,Tolos:2006ny,Tolos:2008di,Cabrera:2014lca}. Finally, we will discuss  the properties of the density and molecular-probability  dependence of the $D_{s0}^\ast(2317)^+/D_{s0}^\ast(2317)^-$  and $D_{s1}(2460)^+/D_{s1}(2460)^-$  signatures in matter, and the possibility of using future measurements of the resulting charge-conjugation asymmetries to infer details on the inner dynamics of these exotic states.

The present manuscript is organized in the following way. In Sec.~\ref{sec:formalism} we present the formalism for $D^{(\ast)} K$ and $\olsi D{}^{(\ast)} \olsi K$ scattering inside a dense medium of nucleons. We start discussing the heavy and light meson spectral functions in Sec.~\ref{sec:SpectralFunctions}, and afterwards we devote Secs.~\ref{sec:vacuum} and~\ref{sec:inmedium} to show $D K$ and $\olsi D \olsi K$ scattering in the vacuum and in a dense nuclear medium, respectively. We proceed by commenting on the situation for $D^\ast K$ and $\olsi D{}^\ast \olsi K$ in Sec.~\ref{sec:Ds1}, followed by Sec.~\ref{sec:results} in which we discuss the obtained results. Lastly, the conclusions are presented in Sec.~\ref{sec:conclusions}.

\section{Formalism} \label{sec:formalism}
In this section we present the main features of the in-medium $S$-wave $D^{(*)} K$ and $\olsi D{}^{(*)} \olsi{K}$ scattering formalism. We closely follow our previous works of Refs.~\cite{Albaladejo:2021cxj,Montesinos:2023qbx}, in which the $D^{\ast}\olsi{D}$, $D\olsi{D}{}^{\ast}$, $D^{\ast}D$ and $\olsi{D}{}^{\ast}\olsi{D}$ channels were explored in order to describe the $X(3872)$ and the $T_{cc}(3875)^+$ states for different molecular-probability scenarios.  However, in this work we do not deal with interactions between charmed mesons, but rather with the Goldstone boson scattering off charmed mesons  and  thus  we will  have to make connection to the interactions deduced in HMChPT.

\subsection{\texorpdfstring{\boldmath $D^{(\ast)}$, $\olsi D{}^{(\ast)}$, $K$ and $\olsi K$}{D and K} spectral functions}\label{sec:SpectralFunctions}

Let us focus first on the $D^{(\ast)}$ and $\olsi D{}^{(\ast)}$ spectral functions ($S_{M=D^{(\ast)},\olsi D{}^{(\ast)}}$), which are determined by their in-medium self-energies ($\Pi_M$). We already employed these spectral functions in our previous analyses  of Refs.~\cite{Albaladejo:2021cxj,Montesinos:2023qbx}. As described in Refs.~\cite{Tolos:2009nn,Garcia-Recio:2010fiq,Garcia-Recio:2011jcj} (see also Ref.~\cite{Tolos:2013gta} for a review), the charmed-meson self-energies are computed following a self-consistent  procedure, which relies on  the vacuum $D^{(\ast)}N$ and $\olsi{D}{}^{(*)}N$ interactions  derived from a $S$-wave effective Lagrangian that \begin{enumerate*}[label=\textit{\roman*)}] \item accounts for the lowest-lying pseudoscalar and vector mesons as well as $1/2^+$ and $3/2^+$ baryons, \item implements heavy quark spin symmetry (HQSS), and \item reduces to the chiral SU(3) Weinberg-Tomozawa interaction term in the sector where only Goldstone bosons are involved~\cite{Garcia-Recio:2008rjt,Gamermann:2010zz,Romanets:2012hm} \end{enumerate*}. The amplitudes obtained from this Lagrangian are used as kernels to  solve the Bethe-Salpeter equation (BSE), which restores elastic unitarity in coupled-channels (see for instance Ref.~\cite{Romanets:2012hm}).  

Once the self-energies are determined, the corresponding spectral functions are obtained as 
\be\label{e:selfenergy}
S_M(E,\, \vec{\,q};\, \rho) = -\frac{1}{\pi} \frac{\mathrm{Im} \left[\Pi_M(E,\, \vec{\,q};\, \rho)\right]}{\left\lvert E^2 - \vec{\,q}^{\,2} - m_M^2 - \Pi_M(E,\, \vec{\,q};\, \rho)\right\rvert^2},
\ee
with $\Pi_M(E,\, \vec{\,q};\, \rho)$  the self-energy of a certain  meson $M$ of mass $m_M$, which depends on its energy ($E$), three-momentum ($\vec{q}$\,) and isospin-symmetric nuclear density $\rho$. Variables are referred to the reference system where the nuclear matter is at rest, and obviously the meson self-energies do not depend on the direction of $\vec{q}$ when they are embedded in an isotropic nuclear environment.   The resulting $D^{(\ast)}$ and $\olsi{D}{}^{(\ast)}$ spectral functions were shown in Fig. 1 of Ref.~\cite{Montesinos:2023qbx}. For increasing densities, one can observe the broadening of the quasi-particle peak and the appearance of other secondary peaks, which correspond to the excitation of different resonance-hole states. Their structure has been discussed in previous works \cite{Montesinos:2023qbx,Albaladejo:2021cxj}, hence for the sake of brevity we will not enter into the details here.

\begin{figure}[t]
    \centering
    \includegraphics[height=.45\textwidth]{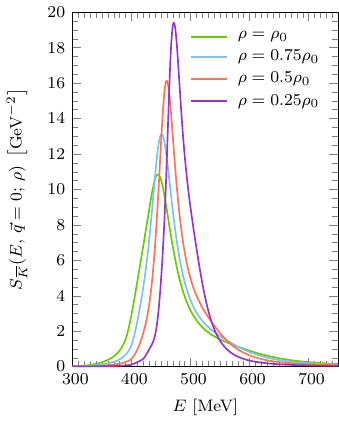}
    \hspace{5mm}
    \includegraphics[height=.45\textwidth]{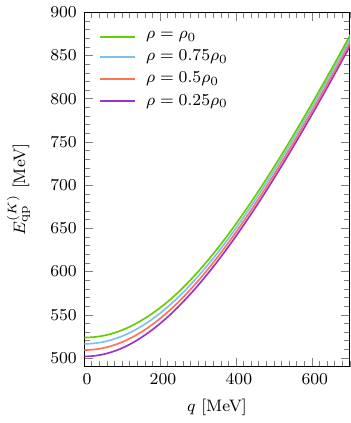}
    \caption{Left:  Energy dependence of the $\olsi K$ spectral function at zero three-momentum ($\vec{q}=0$) for different values of the nuclear density in units of $\rho_0=0.17 \ \text{fm}^{-3}$. Right: $K$ quasi-particle energy (Eq.~\eqref{eq:eqsp}) as a function of the modulus of the kaon three-momentum $q$ ($=|\vec{q}\,|)$ for different densities.}
    \label{fig:KSpectralFunctions}
\end{figure}

Regarding the $K$ and $\olsi K$ spectral functions, a self-consistent chiral unitary approach in coupled channels was used as described in Ref.~\cite{Tolos:2008di}. The computation incorporates the $S$- and $P$-waves of the kaon-nucleon interaction in a self-consistent manner. The in-medium solution accounts for the implementation of Pauli blocking on baryons in the intermediate meson-baryon propagator, the inclusion of the  $K$ and $\olsi K$  self-energies in the $K$ and $\olsi K$ propagation in dense matter, respectively, together with the incorporation of self-energies of all hadrons (pions and baryons) in the intermediate states.

The energy dependence of the $\olsi K$ spectral function is shown in the left plot of Fig.~\ref{fig:KSpectralFunctions} for four different densities and $\vec{q}=0$.
We observe that the $\olsi{K}$ quasiparticle peak energy $E_\mathrm{qp}$, defined as
\be
  E_\mathrm{qp}^2-q^2-m_{\olsi K}^2-\mathrm{Re}\left[\Pi_{\olsi K}(E_\mathrm{qp},\, q;\, \rho)\right] = 0. \label{eq:eqsp}
\ee
with $q$ the modulus of the $\olsi K$ three-momentum, is located at a lower energy than the free $\olsi K$ mass in dense matter. Moreover, the $\olsi K$ falls off slowly on the right-hand side of the quasiparticle peak. This is due to the presence of $\Lambda(1405) N^{-1}$ excitation for energies above the quasiparticle energy. As density increases, the quasiparticle peak gains attraction whereas the spectral function becomes wider due to the dilution of the $\Lambda(1405)$ with density, as thoroughly discussed in Refs.~\cite{Tolos:2006ny,Tolos:2008di,Cabrera:2014lca}.


In sharp contrast, the kaon spectral function has little structure, being akin to a Dirac delta function peaked around the quasi-particle energy $E_\mathrm{qp}$. Indeed, we find that\footnote{This is obtained taking the zero limit for the imaginary part of  the kaon self-energy and approximating by $(2 E_\mathrm{qp})^{-1}$ the quasi-particle strength Jacobian (see for instance Ref.\,\cite{Nieves:2017lij}) 
\begin{equation}
\left |2E-\frac{\partial\mathrm{Re}\left[\Pi_K(E,\, q;\, \rho)\right]}{\partial E} \right|^{-1}_{E=E_\mathrm{qp}}= \frac1{2 E_\mathrm{qp}}    
\end{equation}} 
\be
    S_K(E,\, q ; \, \rho) \approx \frac{\delta\left(E - E_\mathrm{qp}(q;\, \rho)\right)}{2 E_\mathrm{qp}(q;\, \rho)},
\ee
is an excellent approximation for the kaon spectral function, which allows to simplify the numerical computation of the in-medium $D{}^{(*)}K$ loop function.  The in-medium kaon quasi-particle energy is shown in the right plot of Fig.~\ref{fig:KSpectralFunctions} for various nuclear densities as a function of the modulus of the kaon three-momentum. We observe a very mild dependence on the medium density, or in other words, we find very small density corrections to the relativistic dispersion relation. This is expected because of the small $KN$ cross section, as pointed out in the Introduction, and discussed in Ref.~\cite{Tolos:2008di}.

\subsection{\boldmath Vacuum \texorpdfstring{$D K$ and $\olsi{D}  \olsi{K}$}{DK} scattering amplitudes} \label{sec:vacuum}

We start by considering $DK$ elastic $S$-wave scattering in the $I(J^P)=0(0^+)$  sector to dynamically generate the $D_{s0}^\ast(2317)$ out of the unitarity loops. We neglect here explicit coupled-channels effects from the $D_s\eta$, whose threshold is located around $150\,\text{MeV}$ above the $DK$ one, and its mild-energy effects around the $D_{s0}^\ast(2317)$ should be safely accounted for some re-tuning of the LECs (see also Refs.~\cite{Aceti:2014ala,Hyodo:2013nka,Sekihara:2014kya,Albaladejo:2015kea}). As in our previous works of Refs.~\cite{Montesinos:2023qbx,Albaladejo:2021cxj}, we will introduce two families of energy-dependent contact potentials, expanded around threshold:\footnote{ We work in the isospin limit and take $m_{D^+}=m_{D^0}=m_D$ and $m_{K^+}=m_{K^0}=m_K$.}
\begin{subequations}
\label{e:Potential}
\begin{align} 
    V_A(s) &= C_1 + C_2\, [s-(m_D+m_K)^2] , \\
    V_B(s) &= \left(C_1^\prime + C_2^\prime\,  [s-(m_D+m_K)^2]\right)^{-1} ,
\end{align}  
\end{subequations}
where $s=P^2$, with $P^\mu$ the total four-momentum of the $DK$ pair and $C_1^{(\prime)}$ and $C_2^{(\prime)}$ adjustable LECs, which are fixed by imposing that the $T$-matrix presents a pole at the mass ($m_0$) of the $D_{s0}^\ast(2317)$ and that the coupling ($g_0$) of this state to the $DK$ channel is such that gives rise to a molecular probability content $P_0$~\cite{Weinberg:1965zz}, as we will detail below.

We obtain the $T$-matrix from the solution of the BSE, within the on-shell approximation~\cite{Nieves:1999bx},
\be
T^{-1}(s)=V^{-1}(s)-\Sigma_0(s), 
\ee
where $\Sigma_0(s)$ is the two-meson loop function in the vacuum,
\begin{align}
    \Sigma_0(s) = i \int \frac{d^4q}{(2\pi)^4} \Delta_D(P-q)  \Delta_K(q), \\
    \Delta_{M}(q) = \frac{1}{(q^0)^2-\vec{\,q}^{\,2} - m_M^2 + i\varepsilon},
\end{align}
which requires to introduce an ultraviolet (UV) regulator in the $d^3q$ integration to make
the two-point function $\Sigma_0(s)$ finite. In this work, we will use a sharp momentum cutoff, $\Lambda=0.7$ GeV. 

To determine the LECs of the $DK$ potential, we impose  in the first Riemann sheet (FRS) of the BSE amplitude 
\be
\label{e:TmatrixConditions}
T^{-1}(m_0^2) = 0, \qquad \frac{dT^{-1}(s)}{ds}\Big |_{s=m^2_0} = \frac{1}{g^2_0} =-\frac1{P_0}\, \left. \frac{\partial\Sigma_0(s) }{\partial s}\right|_{s=m_0^2} ,
\ee
where  in the last condition,  we have made use of the relation between the molecular content of a bound state, its coupling (residue)  to the two-hadron pair and the derivative of the loop function at the pole position~\cite{Gamermann:2009uq}. Thus, we obtain  expressions for $V_A(s)$ and $V_B(s)$, given in Eqs.~(15) and (16) of Ref.~\cite{Montesinos:2023qbx},\footnote{There is a typo in the second term of the right-hand side of Eq.~(16) of Ref.~\cite{Montesinos:2023qbx} and there it should appear $(s-m_0^2)$ instead of $(s^2-m_0^2)$.} in terms of $m_0$, $\Sigma_0(m_0^2)$ and the derivative $\Sigma^\prime_0(m_0^2)$. Note that the numerical value of $\Sigma_0(m_0^2)$ depends strongly on the UV cutoff $\Lambda$, while  $\Sigma^\prime_0(m_0^2)$ has only a residual dependence. The free-space  $D K$ and $\olsi D \olsi K $  $T$-matrices, $V$-potentials and loop functions $\Sigma_0$  are identical due to the charge-conjugation symmetry.

However, some discussion on the Weinberg compositeness concept and further developments are in order here. In  Ref.~\cite{Weinberg:1965zz} the experimental values for the scattering length ($a$) and effective range ($r$) from $pn$ scattering were used by Weinberg to show evidence that the deuteron is composite. Nevertheless, this does not follow from the evaluation of the so-called compositeness $X$ ($P_0$ throughout the present manuscript)  as $X=1-Z=1/\sqrt{1+2r/a}$, that gives the meaningless result of $X=1.68>1$ for the molecular probability\,\cite{Matuschek:2020gqe, Esposito:2021vhu,Li:2021cue,Song:2022yvz}, as one would naively infer from Ref.~\cite{Weinberg:1965zz}. The key token for the deuteron compositeness is the fact that $r$ is small and positive of the order of the range $\sim m_\pi^{-1}$ of the $pn$ interaction, as indicated by Weinberg, rather than large and negative. Therefore, any conclusion about the nature of an exotic state based uniquely on the computation of $X$ can be misleading, as it neglects ${\cal O}(1/\gamma_b)$ corrections, with $\gamma_b=\sqrt{-2\mu E_b}$ the binding momentum. Here, $\mu$ and $E_b(<0)$ are the reduced mass of the $pn$ pair and the deuteron binding energy ($-2.2\,\text{MeV}$), respectively. Several later works have worked out different applications, re-derivations, re-interpretations and extensions of Weinberg's compositeness relations~\cite{Baru:2003qq, Gamermann:2009uq, Baru:2010ww, Hanhart:2011jz,Aceti:2012dd,Hyodo:2011qc,Sekihara:2014kya,Kamiya:2015aea,Garcia-Recio:2015jsa,Guo:2015daa, Kamiya:2016oao,Sekihara:2016xnq,Oller:2017alp,Matuschek:2020gqe,Esposito:2021vhu,Li:2021cue,Kinugawa:2021ykv,Song:2022yvz,Sazdjian:2022kaf, Albaladejo:2022sux,Song:2023pdq,Dai:2023kwv}, but so far there is no clear consensus on how to apply these relations to determine the compositeness or elementariness of specific states, in particular if they are not bound.
Note that in spite of $Z$ being defined as a bare-state probability in Eq.~(18) of Ref.~\cite{Weinberg:1965zz}, it is not fully an observable as the bare compact QCD states are not physical, and the effects produced by the interacting hadron cloud should be considered. Indeed, $Z$ is a renormalization field factor~\cite{Weinberg:1965zz,Matuschek:2020gqe}, being a scheme and even
regularization dependent quantity. However, in the weak binding limit ($\gamma_b \ll \beta$,  with $1/\beta$ providing an estimate for the interaction range corrections) and for two particle $S$-wave scattering, the quantity $Z$ is dominated by a term obtained from  the residue of the two-hadron scattering amplitude $f(E)$ at the physical pole $E=E_b$~\cite{Weinberg:1965zz,Matuschek:2020gqe}. Given that the latter is the effective coupling of the bound state to the continuum channel, a  measurable quantity, this model-independent contribution to $Z$ becomes a valuable measure of the compositeness.  The scheme and scale dependent terms of $Z$, for instance those analytic in $E$, have to be
fixed by some renormalization condition, but importantly they are suppressed by a factor of the order $\mathcal{O}(\gamma_b/\beta)$~\cite{Weinberg:1965zz,Matuschek:2020gqe}. More specifically, in the weak binding limit
\begin{equation}
X = 1-Z \simeq \frac{\mu \widehat{g}^2}{\gamma_b}+{\cal O}\left(\gamma_b/\beta\right), \qquad \widehat{g}^{\,2}=\lim_{E\to E_b} (E_b-E)f(E)\,. \label{eq:zgt}   
\end{equation}
The above equation shows how the effective coupling $\widehat{g}^{\,2}$, though it does not fully determine the sub-leading  $\mathcal{O}(\gamma_b/\beta)$ contributions to $Z$,  provides most of the molecular probability $X=1-Z$. This will be the scheme followed in this work.\footnote{The relations of Eq.~\eqref{e:TmatrixConditions} are consistent to Eq.~\eqref{eq:zgt}. The couplings $\widehat{g}$ and $g_0$ differ because of the difference of normalization between the scattering amplitudes $f$ and $T$ used in Eqs.~\eqref{eq:zgt}  and \eqref{e:TmatrixConditions}, respectively. In addition for a shallow bound state, the leading term of $\frac{\partial\Sigma_0(s) }{\partial s}\big|_{s=m_0^2}$ is proportional to $1/\gamma_b$.} Further discussions and references can be found in Ref.\,\cite{Albaladejo:2022sux}.

The LO HMChPT $S$-wave isoscalar $DK$ potential reads~\cite{Guo:2009ct}
\begin{eqnarray}
 V_{\chi LO}(s)&=& \frac{-3s+2m_K^2+2m_D^2+(m_D^2-m_K^2)^2/s}{4f^2} ,
\end{eqnarray}
with $f\sim 93\,\text{MeV}$. In the vicinity of the position of the $D_{s0}^\ast(2317)$, this potential admits an expansion of the  $V_A-$ or $V_B-$types  introduced in Eq.~\eqref{e:Potential}.  This LO interaction was used in Ref.~\cite{Albaladejo:2016hae} to relate the $D_{s0}^\ast(2317)$ state to  structures right above threshold seen in the experimental $D^0 K^+$ and $\olsi D{}^0 K^-$ invariant mass spectra of the BaBar reactions $B^+\to \olsi D{}^0 D^0K^+$ and $B^0\to D^-D^0K^+$  \cite{BaBar:2014jjr} and the LHCb $B_s\to \pi^+ \olsi D{}^0 K^-$  one  \cite{LHCb:2014ioa}. The analysis carried out in Ref.\,\cite{Albaladejo:2016hae} found a  pole  in $DK$ amplitude at a mass of $2315\pm 17\,\text{MeV}$, with a molecular probability of $70^{+6}_{-10}$\%. In that work the LO HMChPT was unitarized using  the on-shell BSE, which was renormalized by means of a subtraction constant which was fitted  to the experimental LHCb and BaBar distributions. Here, we use instead a sharp-cutoff UV regulator, because it is more appropriate in order to incorporate nuclear medium effects (see discussion on the last paragraph of Sec.\ref{sec:inmedium}). Both renormalization schemes are equivalent, and the results of \cite{Albaladejo:2016hae} are fairly well reproduced in the region of interest using $\Lambda= 875\pm 85\,\text{MeV}$ (see Eq.~(52) of Ref.~\cite{Garcia-Recio:2010enl}).

\begin{figure}[t]
    \centering
    \includegraphics[width=.45\textwidth]{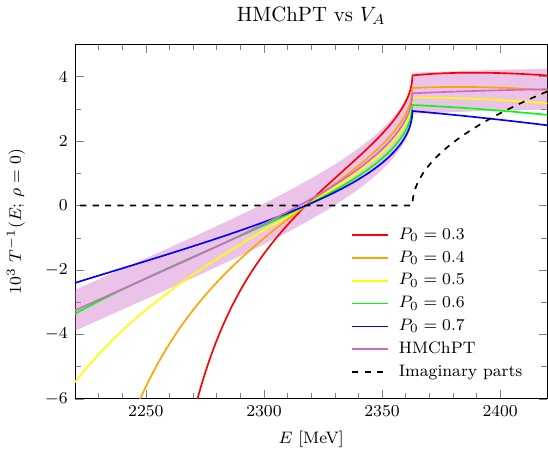}
    \hspace{5mm}
    \includegraphics[width=.45\textwidth]{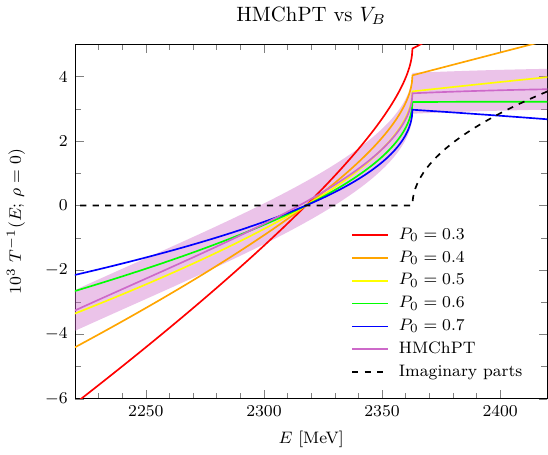}
    \caption{Real parts of the free-space $S$-wave isoscalar $DK$ inverse amplitudes obtained using the LO HMChPT scheme followed in  Ref.~\cite{Albaladejo:2016hae} (magenta band) and the $V_A$ (left panel) and $V_B$ (right panel)  families of potentials (Eq.~\eqref{e:Potential}), adjusted for different $D_{s0}^\ast(2317)$ molecular probabilities ($P_0$), as  functions of the center of mass energy ($E$) of the $DK$ pair. The error band of the HMChPT result accounts  for the uncertainty on the subtraction constant fitted in Ref.~\cite{Albaladejo:2016hae} to the combined BaBar and LHCb mass distributions. The imaginary part (black dashed line)  of the inverse amplitude for real $E$ ($=\sqrt{s}$) is the same independently of the potential and of the used regularization method, as it is derived from unitarity, i.e.  Im[$T^{-1}(s)]=-$Im[$\Sigma_0(s)]= H\left(s-(m_D+m_K)^2\right)\lambda^{1/2}(s,m_D^2,m_K^2)/(16\pi s)$, with $H$ and $\lambda$ the step and K\"allen functions, respectively.}
    \label{fig:PotentialComparison}
\end{figure}

In Fig.~\ref{fig:PotentialComparison}, we compare the real parts of $T^{-1}$ obtained using $V_A$ and $V_B$ families of potentials, for several molecular probabilities,  and that of the LO HMChPT scheme of Ref.~\cite{Albaladejo:2016hae}.  We see that the LO HMChPT result in the region of interest around the $D_{s0}^\ast(2317)$ mass ($2280\,\text{MeV} < E < 2390\,\text{MeV}$)  is reasonably well reproduced using both the $V_A$ and $V_B$ families of potentials and  molecular probabilities between $0.5$ and $0.7$. Furthermore, if we pay attention to the $V_A$ and $V_B$ potential families, we observe that they are extremely similar for high values of the molecular probability (see  for instance $P_0=0.7$), while some differences arise when small values of $P_0$ are considered.\footnote{Actually, as was discussed in Ref.~\cite{Albaladejo:2021cxj}, in the $P_0\to 1$ scenario both potentials become equal and energy-independent $(V_A(s)=V_B(s)=1/\Sigma_0(m_0)$, while in the $P_0\to0$ limit, both potentials become ill-defined. This is because in the pure non-molecular case, $P_0=0$, the state does not couple to the two-meson channel.} 

\subsection{\boldmath \texorpdfstring{$D K$ and $\olsi{D} \olsi{K}$}{DK} scattering in isospin-symmetric nuclear matter} \label{sec:inmedium}

For simplicity, we will restrict our analysis to the isospin limit (as we mentioned in Sec.~\ref{sec:vacuum}), focusing solely on the modifications of the $T$ amplitude  caused by the changes of the $D K$ and $\olsi D \olsi K$ loop functions, $\Sigma(s;\rho)$ and $\ols \Sigma(s;\rho)$, respectively, when they are calculated in a nuclear environment.   The medium modifications are induced by the meson self energies $\Pi_M(q^0,\vec{q}\,;\,\rho)$ arising from interactions between $K$, $\olsi K$, $D$ and $\olsi{D}$ mesons and the nucleons of the medium. While, by construction,  these self-energies vanish in the vacuum ($\rho = 0$), they produce substantial modifications in the dispersion relations of the mesons embedded in nuclear matter. As mentioned in the Introduction, the dense nuclear medium breaks charge-conjugation symmetry and as a consequence,  $D K$ and $\olsi D \olsi K$ scattering amplitudes will no longer be the same. We expect large asymmetries since, as we have seen in Sec.~\ref{sec:SpectralFunctions}, the spectral functions for the mesons and their corresponding anti-particles are radically different.

The K\"allen–Lehmann representation of the meson propagators
\be\label{eq:dressed-pro}
 \Delta_M(q;\rho)  = \int_0^\infty d\omega \left( 
 \frac{S_M(\omega,\lvert \vec{q}\, \rvert; \rho)}{q^0 - \omega + i\varepsilon} - 
 \frac{S_{\olsi{M}}(\omega,\lvert \vec{q}\, \rvert;\rho)}{q^0 + \omega - i\varepsilon} 
 \right) \ ,
\ee
allows us to rewrite the in-medium two-meson loop functions as~\cite{Albaladejo:2021cxj}
\begin{subequations}%
\label{eq:SigmaAndSigmabar}
\begin{align}
    \Sigma(s=E^2;\rho) 
    &= \frac{1}{2\pi^2}\left[\mathcal{P}\!\!\!\int_0^\infty d\Omega \left( \frac{f_{DK}(\Omega;\rho)}{E-\Omega} - \frac{f_{\olsi{D}\olsi{K}}(\Omega;\rho)}{E+\Omega} \right) - i\pi f_{DK}(E;\rho) \right]\label{eq:DstarDLoopfunction} \ , \\
 \ols{\Sigma}(s=E^2;\rho) 
    &= \frac{1}{2\pi^2}\left[\mathcal{P}\!\!\!\int_0^\infty d\Omega \left( \frac{f_{\olsi{D}\olsi{K}}(\Omega;\rho)}{E-\Omega} - \frac{f_{DK}(\Omega;\rho)}{E+\Omega} \right) - i\pi f_{\olsi{D}\olsi{K}}(E;\rho) \right] \ ,\label{eq:barDstarDLoopfunction}
\end{align}
\end{subequations}
where the $\mathcal{P}$ symbol stands for the Cauchy principal value of the integral, and where the auxiliary $f_{UW}$ function is defined as
\begin{equation}
    f_{UW}(\Omega;\rho) = \int_0^\Lambda dq\, q^2 \int_0^\Omega d\omega \ S_U \left( \omega, |\vec{q}\,|;\rho\right) S_W \left(\Omega-\omega,|\vec{q}\, |;\rho\right)\label{eq:deff}.
\end{equation}
 In Eq.~\eqref{eq:deff} we have incorporated the sharp cutoff $\Lambda=0.7$ GeV, employed in this work,  in the momentum integral to manage the UV divergence. Note that the spectral functions depend on $q^0$ and on the magnitude of $\vec{q}$, but not on any specific direction, when considering spherically symmetric nuclear matter. Furthermore, we have assumed in derivation of Eq.~\eqref{eq:deff} that the center of mass of the meson pair is also at rest, $\vec{P}=0$, thus resulting in $P^2 = ({P^0})^2 = s$. 
 
 Using the Dirac delta approximation for the kaon spectral function, we can further simplify the expression for the auxiliary function $f_{DK}$, yielding:
\be
f_{DK}(\Omega;\rho) = \int_0^\Lambda dq\, q^2  \frac{S_D (\Omega-E_\mathrm{qp}^{(K)},\, |\vec{q}\,|;\rho)}{2E_\mathrm{qp}^{(K)}} \ .
\ee
Finally, we obtain the $DK$ amplitude $T^{-1}(s;\rho)$ in the nuclear medium of density $\rho$ as  
\begin{subequations}%
\label{eq:Vmenosunoeff}%
\begin{align}
T^{-1}(s;\rho) & = V_{\rm eff}^{-1}(s;\rho) -\Sigma_0(s)\,,\\
V_{\rm eff}^{-1}(s;\rho) & = V^{-1}(s)+\delta\Sigma(s;\rho)\,,\\
\delta\Sigma(s;\rho) & =\Sigma_0(s)-\Sigma(s;\rho)\,,
\end{align}
\end{subequations}
where $V_\mathrm{eff}$ includes the effects of the nuclear medium, and its density behavior will allow us to discuss how the nuclear environment effectively changes the interaction between the two mesons. In the $\olsi D \olsi K$ case, we analogously define
\begin{subequations}%
\label{eq:Vbareff}%
\begin{align}
\ols{V}_{\rm eff}^{-1}(s;\rho) & = V^{-1}(s)+\delta\ols{\Sigma}(s;\rho)\,,\\
\delta\ols{\Sigma}(s;\rho) & =\Sigma_0(s)-\ols{\Sigma}(s;\rho)\,,
\end{align}
\end{subequations}
and hence
\begin{equation}
    \ols{V}_{\rm eff}^{-1}(s;\rho)- V_{\rm eff}^{-1}(s;\rho) = \Sigma(s;\rho)-\ols{\Sigma}(s;\rho)\,. \label{eq:diff-veff} 
\end{equation}

Before finalizing this section, a discussion about the use of the on-shell BSE in nuclear matter is in order here. The on-shell BSE may not work as well when considering the effects of matter. Aware of this problem, in this work we have replaced the dimensional regularization scheme adopted in the original HMChPT NLO works of Refs.~\cite{Guo:2009ct,Albaladejo:2016lbb} to compute the free-space two-meson loop functions by the use of a sharp cutoff in the nuclear medium. This is consistent with our previous calculations of the $D^{(*)}, \olsi D{}^{(*)}$ and $\olsi K$ spectral functions in a nuclear environment~\cite{Tolos:2009nn,Garcia-Recio:2010fiq,Garcia-Recio:2011jcj,Tolos:2008di}. Moreover the antikaon-nucleus optical potential calculated in this way (on-shell BSE and a sharp cutoff to compute the loop-function)~\cite{Ramos:1999ku, Tolos:2008di} leads to an excellent description of the kaonic atom data~\cite{Baca:2000ic,Hirenzaki:2000da}.  Our work yields results with regard to the charge-conjugation asymmetry effect in the nuclear medium and the line shape sensitivity to the molecular probability, and the consideration of off-shell terms are not going to qualitatively modify the conclusions of our present investigation.

\subsection{\boldmath \texorpdfstring{$D^\ast K$ and $\olsi{D}{}^\ast \olsi{K}$}{D*K} scattering amplitudes and the \texorpdfstring{$D_{s1}(2460)$}{Ds1(2460)}}\label{sec:Ds1}

HQSS is an approximate symmetry of QCD that renders the QCD lagrangian independent of the quark spin when heavy quarks are involved. This gives rise to approximate degenerate doublets of spin partners, like the $D$ and $D^\ast$ mesons. The mass gap between these two latter mesons correspond approximately to the mass of the pion. The isoscalar axial  ($J^P=1^+$) $D_{s1}(2460)$ state also lies at an energy of about one pion mass above that of the isoscalar scalar  ($J^P=0^+$) $D_{s0}^\ast(2317)$, and it is commonly accepted that  this pair of mesons form a HQSS doublet, where the light degrees of freedom are coupled to isospin zero and spin-parity $1/2^+$ (see for example the discussion in Ref.~\cite{Albaladejo:2018mhb}).

Within our formalism, we consider the axial $D_{s1}(2460)^+$ as a dynamically  state generated by the isoscalar $S$-wave $D^\ast K$ scattering. In addition, because of  HQSS, the $I(J^P)=0(0^+)$ $D K$ and $I(J^P)=0(1^+)$ $D^\ast K$  amplitudes will be the same, replacing the mass of the pseudoscalar meson $D$ by that of the vector meson $D^*$,  up to very small HQSS breaking effects in the coefficients $C_{1,2}^{(\prime)}$ of the $V_A$ and $V_B$ potentials [Eq.~\eqref{e:TmatrixConditions}] induced by the difference between the $[m_{D_{s1}(2460)}-m_{D_{s0}^\ast(2317)}]$  and the $[m_{D^*}-m_D]$ mass splittings. 

Nuclear matter density effects are then incorporated through the in-medium $D^* K$ (and $\olsi{D}{}^\ast \olsi{K}$) loop functions, which are computed in the same way as was presented in Sec.~\ref{sec:inmedium} from the meson spectral functions.

\section{Nuclear medium results} \label{sec:results}
In this section we present nuclear medium results  for the $D^{(*)}K$ and $\olsi D{}^{(*)} \olsi K$ loop functions and the modulus squared of the $T$-matrix for  the $I(J^P)=0(0^+)$ and $I(J^P)=0(1^+)$  $D^{(*)}K$ and $\olsi D{}^{(*)} \olsi K$ channels, where the $D_{s0}^*(2317)$ and  $D_{s1}^*(2460)$ poles show up in the vacuum.

\begin{figure}[t]
    \centering
    \includegraphics[width=.45\textwidth]{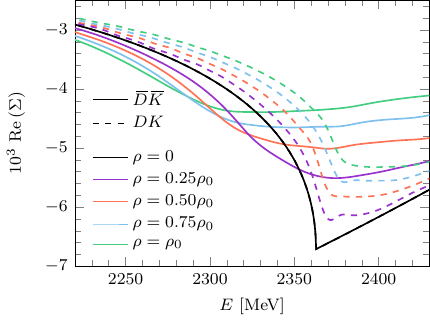}
    \hspace{5mm}
    \includegraphics[width=.45\textwidth]{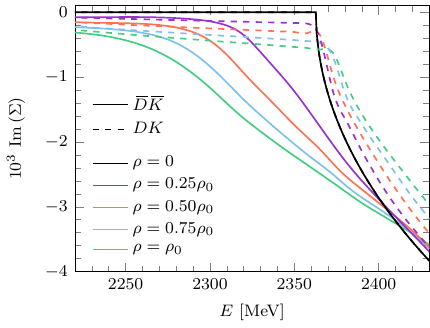}
    \caption{Real (left) and imaginary (right) parts of the $\olsi{D}  \olsi{K}$ (solid lines) and $DK$ (dashed lines) loop functions. We show results for different values of the nuclear medium density (in units of $\rho_0=0.17 \ \text{fm}^{-3}$) as a function of the center of mass energy of the heavy light-Goldstone meson pair.}
    \label{fig:barLoopComparison}
\end{figure}

In  Fig.~\ref{fig:barLoopComparison} we compare the lineshapes of the $\olsi D \olsi K$ (solid) and the $DK$ (dashed) loop functions for different values of the nuclear density, ranging from $0$ to $\rho_0=0.17 \ \text{fm}^{-3}$. Both loop functions coincide in the vacuum, as imposed by charge-conjugation symmetry. However, we observe that both real (left plot) and imaginary parts (right plot) of $\Sigma(s\,;\,\rho)$ and $\ols{\Sigma}(s\,;\,\rho)$ significantly deviate as the density increases. The charge-conjugation asymmetry pattern  found here is much more pronounced than the $D^\ast D$ \textit{versus} $\olsi D{}^\ast \olsi D$ one found in Ref.~\cite{Montesinos:2023qbx}, in the context of the study of the $T_{cc}(3875)^+$ and $T_{\olsi c \olsi c}(3875)^-$ tetraquarks embedded in a nuclear environment. 

Analyzing the real and imaginary parts in more depth we see that, on the one hand, the imaginary part of the $\olsi D \olsi K$ loop function is notably bigger (in absolute value) than that of the $D K$ around $E=2320$ MeV, which is the region where the $D_{s0}^\ast$ pole appears in the vacuum. Hence, the $D_{s0}^{\ast -}$  in the nuclear medium  would have a larger width than  the vacuum charge-conjugate partner $D_{s0}^{\ast +}$. This is due to the sizable broadening  of the quasi-particle peak for the antikaon in the medium (see Fig.~\ref{fig:KSpectralFunctions}),  in sharp contrast to the Dirac delta peak found, in a very good approximation, for the kaon. On the other hand, we see that around the $D_{s0}^\ast$ vacuum mass, the real part of the $\olsi D \olsi K$ loop function is significantly more negative than that of the $D K$ one. Through Eq.~\eqref{eq:diff-veff}, we get that in this region of energies
\be\label{eq:ReVeffComp}
\mathrm{Re}\left(\ols{V}_{\rm eff}^{-1}\right) - \mathrm{Re}\left(V_{\rm eff}^{-1}\right) > 0 ,
\ee
and hence, if we ignore the imaginary part of the effective potential, Eq.~\eqref{eq:ReVeffComp} would imply that $V_{\rm eff}$ is greater than $\ols{V}_{\rm eff}$ and therefore more repulsive. This means that the $D_{s0}^{\ast+}$ pole would shift towards higher energies as compared with the $D_{s0}^{\ast-}$ pole, as the $DK$ interaction in the medium would become less attractive as compared with the $\olsi D \olsi K$ one. However, we should point out that ignoring the imaginary part in the effective potential is an approximation, specially for the $\olsi D \olsi K$ case, where it is certainly  not negligible.

For the $D^\ast K$ and the $\olsi D{}^\ast \olsi K$ loop functions, the density patterns are very similar  to the $D K$ and the $\olsi D{}\olsi K$ ones, and hence we do not show them here. We could extract the same conclusions as for the $D K$ and the $\olsi D \olsi K$ cases. The most notable differences are the energy shift of the vacuum threshold, which now moves around the $(m_{D^*}+m_K)$, and the appearance of a larger imaginary part in the medium due to the slightly stronger $D^\ast N$ and $\olsi{D}{}^\ast N$ interactions. However, the overall properties of the loop functions are still dominated by the kaon and antikaon spectral functions.

\begin{figure}[t]
    \centering
    \includegraphics[height=.35\textwidth]{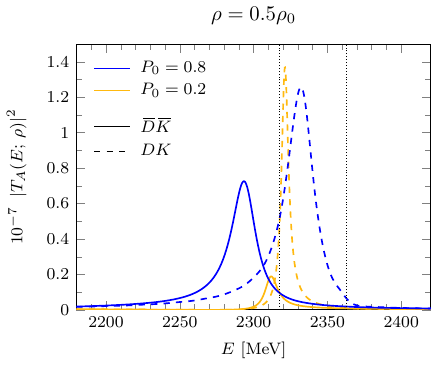}
    \hspace{5mm}
    \includegraphics[height=.35\textwidth]{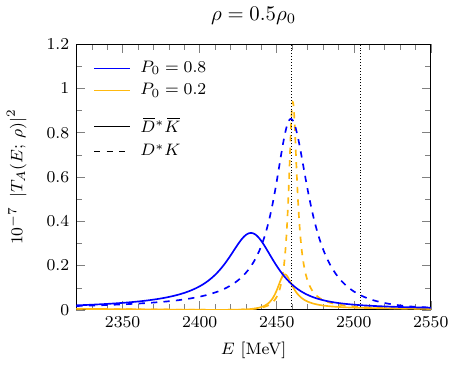}\\
    \vspace{5mm}
    \includegraphics[height=.35\textwidth]{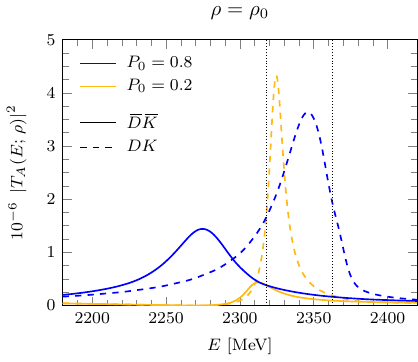}
    \hspace{5mm}
    \includegraphics[height=.35\textwidth]{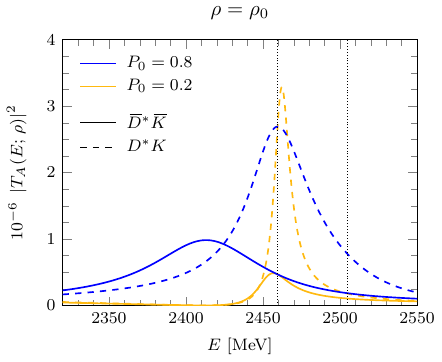}
    \caption{Left panels: In-medium $\olsi{D}\olsi{K}$ (solid lines) and $DK$ (dashed lines) modulus squared amplitudes obtained by solving the BSE using the $V_A(s)$ potential, for vacuum molecular probabilities $P_0=0.2$ (orange) and $P_0=0.8$ (blue), and for nuclear densities $\rho=0.5\rho_0$ (top) and $\rho=\rho_0$ (bottom).  Right panels: Same as left panels but  for $\olsi{D}{}^\ast\olsi{K}$ (solid lines) and $D^\ast K$ (dashed lines)  modulus square amplitudes. In all plots the dotted vertical lines correspond, from left to right, to the vacuum  $D_{s0}^\ast(2317)^\pm$ or $D_{s1}(2460)^\pm$ mass and  $D^{(*)} K$ (and $\olsi D{}^{(*)} \olsi K$) threshold.
    }
    \label{fig:TABbarDsP0Comparison}
\end{figure}

We turn now our attention to the in-medium amplitudes of the $D{}^{(\ast)} K$ and $\olsi{D}{}^{(\ast)} \olsi{K}$ channels. In Fig.~\ref{fig:TABbarDsP0Comparison}, we show the modulus square of these amplitudes using the $V_A$ family of potentials, and two different molecular probabilities ($P_0=0.2$ and $P_0=0.8$) for the $D_{s0}^*(2317)$ and  $D_{s1}^*(2460)$ states in vacuum. These represent two quite opposite scenarios, and the one with higher molecular probability would roughly correspond to that found employing HMChPT in Ref.\,\cite{Albaladejo:2016hae}, as discussed in Subsec.\,\ref{sec:vacuum}. In addition, we have considered two different densities ($\rho=0.5\rho_0$ in the upper plots and $\rho=\rho_0$ in the bottom ones). Results using the $V_B$ family of interactions are very similar for the two molecular contents depicted in the figure, with small differences in the tails of the resonance-peak structures.\footnote{The results obtained with $V_A$ and $V_B$ potentials are more similar among them here than in the $X(3872)$ \cite{Albaladejo:2021cxj} or $T_{cc}(3875)^+$ \cite{Montesinos:2023qbx} cases because of the much larger binding energy of the $D^{\ast}_{s0}(2317)^{\pm}$ state.}

In the left-column plots of Fig.~\ref{fig:TABbarDsP0Comparison} we present results for the isoscalar-scalar  [$I(J^P)=0(0^+)$] $DK$ and $\olsi D \olsi K$ channels. The in-medium $D_{s0}^{\ast +}$ and $D_{s0}^{\ast -}$ lineshapes are displayed by  dashed and solid curves respectively. We see that the $D_{s0}^{\ast +}$ and $D_{s0}^{\ast -}$, which were bound states in the vacuum, acquire some width in the medium. The broadening of the states in matter is more pronounced for the higher molecular component scenario shown in the plots. Furthermore, we see that, as the density and $P_0$ increase, the $D_{s0}^{\ast +}$ [$D_{s0}^{\ast -}$] peak significantly moves towards higher (lower) energies. This behavior was previously pointed out,  when the real part of the effective potential was discussed, and confirmed here accounting also for the  effects induced by the imaginary parts of the loop functions.  The $D_{s0}^{\ast -}$ resonance develops larger widths than the $D_{s0}^{\ast +}$ for all molecular probability and density scenarios considered. This distinctive pattern is largely driven by the quite different renormalization of kaons and antikaons inside the nuclear medium.

The situation for the $D_{s1}^{ +}$ and $D_{s1}^{ -}$ (isoscalar-axial  [$I(J^P)=0(1^+)$] $D^* K$ and $\olsi{D}{}^* \olsi K$ scattering) presented in the right-column plots of Fig.~\ref{fig:TABbarDsP0Comparison} is very similar. However, some differences arise. Mainly, both states get a larger width as compared with their scalar partners. This is because, within the model of Refs.~\cite{Garcia-Recio:2008rjt, Romanets:2012hm} employed here, the $D^{\ast}$ and $\olsi D{}^{\ast}$ interactions with nucleons are slightly stronger than those of the $D$ and $\olsi D$ mesons, as we have already mentioned. In addition, the $D_{s1}^{ +}$ peak is not shifted towards higher energies as was the case for the $D_{s0}^{\ast +}$. These two features come together and make the $D_{s1}^{ +}$ and $D_{s1}^{ -}$ lineshapes in the nuclear medium less distinguishable when compared with the $D_{s0}^{\ast +}$ and $D_{s0}^{\ast -}$ ones, since the differences between the latter ones are more appreciable for any density-molecular component scenario.

\section{Conclusions} \label{sec:conclusions}

We have studied the modifications that a dense nuclear medium produces in the  isoscalar $D^{(*)}K$ and $\olsi{D}{}^{(*)}\olsi{K}$ $S$-wave scattering amplitudes. We have used vacuum effective interactions which  dynamically generate the $D_{s0}^\ast(2317)^\pm$ and the $D_{s1}(2460)^\pm$ bound states, with  different Weinberg compositeness probabilities.   Matter effects are incorporated through the two-meson loop functions,  taking into account the self energies that the $D^{(*)}, \olsi{D}{}^{(*)}, K$ and $\olsi{K}$ develop when embedded in a nuclear medium.

Particle-antiparticle [$D^{(\ast)}_{s0,s1}(2317,2460)^+$ \textit{versus} $D^{(\ast)}_{s0,s1}(2317,2460)^-$] lineshapes are necessarily the same in free space, but we have found extremely different density patterns in matter, arguing that this large charge-conjugation asymmetry mainly stems from the very different kaon and antikaon interactions with the nucleons of the dense medium. Indeed, medium effects violating charge-conjugation symmetry here are larger than those reported in  Ref.~\cite{Montesinos:2023qbx} for $D^\ast D$ and  $\olsi D{}^\ast \olsi D$, in the context of the study of the $T_{cc}(3875)^+$ and $T_{\olsi c \olsi c}(3875)^-$ tetraquarks embedded in a nuclear environment. As in this previous work, we have also seen here that the in-medium spectral functions found for the $D^{(\ast)}_{s0,s1}(2317,2460)^{\pm}$ states  strongly depend on their $D^{(*)}K$/$\olsi{D}{}^{(*)}\olsi{K}$ molecular contents.

With increasing densities and molecular probabilities, we have found that the $D_{s0}^\ast(2317)^+$ peak shifts towards higher energies and becomes less broad than its charge-conjugation partner $D_{s0}^\ast(2317)^-$, whose wider Breit-Wigner-like shape moves more noticeably at lower energies. At half normal nuclear matter density, the change is already so drastic for high molecular component scenarios that the $D_{s0}^\ast(2317)^+$  and $D_{s0}^\ast(2317)^-$ lineshapes hardly overlap.

For the HQSS partners, the axial $D_{s1}(2460)^+$ and $D_{s1}(2460)^-$,  we have found a situation quite similar to the one discussed for the $D_{s0}^\ast(2317)$. However, the $D_{s1}(2460)^{\pm}$ resonant shapes become broader because, within the model of Refs.~\cite{Garcia-Recio:2008rjt, Romanets:2012hm} employed here, the $D^\ast N$ and $\olsi{D}{}^\ast N$ interactions are stronger than the $D N$ and $\olsi D{ N}$ ones. This widening of the distributions produces  that they become slightly less distinguishable. However, the differences between both charge-conjugate channels are still very notable, as the kaon and antikaon spectral functions largely dominate the different $D_{s1}(2460)^+$ and $D_{s1}(2460)^-$ density patterns.

In summary, we have shown that the study of the in-medium behavior of the $D_{s0}^\ast(2317)^\pm$ and $D_{s1}(2460)^\pm$ is a prominent test of their internal structure, since the behavior turns out to be very sensitive to their hadron-molecular content. The presence of nuclear matter breaks charge-conjugation symmetry, and induces different particle-antiparticle lineshapes when these exotic states are produced inside a nuclear environment. If these distinctive density dependencies were experimentally confirmed, it would give support to the presence of important molecular components in these exotic states. This is because if these states were mostly compact four-quark structures rather than molecular-like ones, the density behavior of their in-medium lineshapes, while certainly different, would likely not follow the same patterns found in this work for molecular scenarios.

Another interesting aspect of this study is that it might allow to have a complementary experimental access to the kaon and antikaon self-energies in the nuclear medium, as well as those of the charmed mesons. This would in  turn lead to more experimentally-driven analysis, which could improve on the predictions in this work.

\acknowledgments
This work was supported under contracts No.\, PID2019-110165GB-I00, No.\, PID2020-112777GB-I00, and No.\, PID2022-139427NB-I00 financed by the Spanish MCIN/AEI/10.13039/501100011033/FEDER, UE,  by Generalitat Valenciana under contract PROMETEO/2020/023, and from the project CEX2020-001058-M Unidad de Excelencia ``Mar\'{\i}a de Maeztu''). This project has received funding from the European Union Horizon 2020 research and innovation programme under the program H2020-INFRAIA-2018-1, grant agreement No.\,824093 of the STRONG-2020 project.  M.\,A. and V.\,M.~are supported through Generalitat Valenciana (GVA) Grants No.\,CIDEGENT/2020/002 and ACIF/2021/290, respectively. 
L.\,T. also acknowledges support from the CRC-TR 211 'Strong-interaction matter under extreme conditions'- project Nr. 315477589 - TRR 211 and from the Generalitat de Catalunya under contract 2021 SGR 00171.

\bibliographystyle{JHEP}
\bibliography{references.bib}

\providecommand{\href}[2]{#2}\begingroup\raggedright\begin{thebibliography}{100}

\bibitem{BaBar:2003oey}
{\scshape BaBar} collaboration, \emph{{Observation of a narrow meson decaying
  to $D_s^+ \pi^0$ at a mass of 2.32-GeV/c$^2$}},
  \href{https://doi.org/10.1103/PhysRevLett.90.242001}{\emph{Phys. Rev. Lett.}
  {\bfseries 90} (2003) 242001}
  [\href{https://arxiv.org/abs/hep-ex/0304021}{{\ttfamily hep-ex/0304021}}].

\bibitem{CLEO:2003ggt}
{\scshape CLEO} collaboration, \emph{{Observation of a narrow resonance of mass
  2.46-GeV/c**2 decaying to D*+(s) pi0 and confirmation of the D*(sJ)(2317)
  state}}, \href{https://doi.org/10.1103/PhysRevD.68.032002}{\emph{Phys. Rev.
  D} {\bfseries 68} (2003) 032002}
  [\href{https://arxiv.org/abs/hep-ex/0305100}{{\ttfamily hep-ex/0305100}}].

\bibitem{Godfrey:1985xj}
S.~Godfrey and N.~Isgur, \emph{{Mesons in a Relativized Quark Model with
  Chromodynamics}}, \href{https://doi.org/10.1103/PhysRevD.32.189}{\emph{Phys.
  Rev. D} {\bfseries 32} (1985) 189}.

\bibitem{Godfrey:1986wj}
S.~Godfrey and R.~Kokoski, \emph{{The Properties of p Wave Mesons with One
  Heavy Quark}}, \href{https://doi.org/10.1103/PhysRevD.43.1679}{\emph{Phys.
  Rev. D} {\bfseries 43} (1991) 1679}.

\bibitem{Zeng:1994vj}
J.~Zeng, J.W.~Van~Orden and W.~Roberts, \emph{{Heavy mesons in a relativistic
  model}}, \href{https://doi.org/10.1103/PhysRevD.52.5229}{\emph{Phys. Rev. D}
  {\bfseries 52} (1995) 5229}
  [\href{https://arxiv.org/abs/hep-ph/9412269}{{\ttfamily hep-ph/9412269}}].

\bibitem{Gupta:1994mw}
S.N.~Gupta and J.M.~Johnson, \emph{{Quantum chromodynamic potential model for
  light heavy quarkonia and the heavy quark effective theory}},
  \href{https://doi.org/10.1103/PhysRevD.51.168}{\emph{Phys. Rev. D} {\bfseries
  51} (1995) 168} [\href{https://arxiv.org/abs/hep-ph/9409432}{{\ttfamily
  hep-ph/9409432}}].

\bibitem{Ebert:1997nk}
D.~Ebert, V.O.~Galkin and R.N.~Faustov, \emph{{Mass spectrum of orbitally and
  radially excited heavy - light mesons in the relativistic quark model}},
  \href{https://doi.org/10.1103/PhysRevD.59.019902}{\emph{Phys. Rev. D}
  {\bfseries 57} (1998) 5663}
  [\href{https://arxiv.org/abs/hep-ph/9712318}{{\ttfamily hep-ph/9712318}}].

\bibitem{Lahde:1999ih}
T.A.~Lahde, C.J.~Nyfalt and D.O.~Riska, \emph{{Spectra and M1 decay widths of
  heavy light mesons}},
  \href{https://doi.org/10.1016/S0375-9474(00)00154-8}{\emph{Nucl. Phys. A}
  {\bfseries 674} (2000) 141}
  [\href{https://arxiv.org/abs/hep-ph/9908485}{{\ttfamily hep-ph/9908485}}].

\bibitem{DiPierro:2001dwf}
M.~Di~Pierro and E.~Eichten, \emph{{Excited Heavy - Light Systems and Hadronic
  Transitions}}, \href{https://doi.org/10.1103/PhysRevD.64.114004}{\emph{Phys.
  Rev. D} {\bfseries 64} (2001) 114004}
  [\href{https://arxiv.org/abs/hep-ph/0104208}{{\ttfamily hep-ph/0104208}}].

\bibitem{ParticleDataGroup:2022pth}
{\scshape Particle Data Group} collaboration, \emph{{Review of Particle
  Physics}}, \href{https://doi.org/10.1093/ptep/ptac097}{\emph{PTEP} {\bfseries
  2022} (2022) 083C01}.

\bibitem{Du:2017zvv}
M.-L.~Du, M.~Albaladejo, P.~Fern\'andez-Soler, F.-K.~Guo, C.~Hanhart,
  U.-G.~Mei\ss{}ner et~al., \emph{{Towards a new paradigm for heavy-light meson
  spectroscopy}}, \href{https://doi.org/10.1103/PhysRevD.98.094018}{\emph{Phys.
  Rev. D} {\bfseries 98} (2018) 094018}
  [\href{https://arxiv.org/abs/1712.07957}{{\ttfamily 1712.07957}}].

\bibitem{Weinberg:1965zz}
S.~Weinberg, \emph{{Evidence That the Deuteron Is Not an Elementary Particle}},
  \href{https://doi.org/10.1103/PhysRev.137.B672}{\emph{Phys. Rev.} {\bfseries
  137} (1965) B672}.

\bibitem{Albaladejo:2022sux}
M.~Albaladejo and J.~Nieves, \emph{{Compositeness of S-wave weakly-bound states
  from next-to-leading order Weinberg\textquoteright{}s relations}},
  \href{https://doi.org/10.1140/epjc/s10052-022-10695-1}{\emph{Eur. Phys. J. C}
  {\bfseries 82} (2022) 724}
  [\href{https://arxiv.org/abs/2203.04864}{{\ttfamily 2203.04864}}].

\bibitem{Albaladejo:2016lbb}
M.~Albaladejo, P.~Fernandez-Soler, F.-K.~Guo and J.~Nieves, \emph{{Two-pole
  structure of the $D^\ast_0(2400)$}},
  \href{https://doi.org/10.1016/j.physletb.2017.02.036}{\emph{Phys. Lett. B}
  {\bfseries 767} (2017) 465}
  [\href{https://arxiv.org/abs/1610.06727}{{\ttfamily 1610.06727}}].

\bibitem{Wise:1992hn}
M.B.~Wise, \emph{{Chiral perturbation theory for hadrons containing a heavy
  quark}}, \href{https://doi.org/10.1103/PhysRevD.45.R2188}{\emph{Phys. Rev. D}
  {\bfseries 45} (1992) R2188}.

\bibitem{manohar_wise_2000}
A.V.~Manohar and M.B.~Wise, \emph{{Heavy Quark Physics}}, Cambridge Monographs
  on Particle Physics, Nuclear Physics and Cosmology, Cambridge University
  Press (2000),
  \href{https://doi.org/10.1017/CBO9780511529351}{10.1017/CBO9780511529351}.

\bibitem{Guo:2008gp}
F.-K.~Guo, C.~Hanhart, S.~Krewald and U.-G.~Meissner, \emph{{Subleading
  contributions to the width of the D*(s0)(2317)}},
  \href{https://doi.org/10.1016/j.physletb.2008.07.060}{\emph{Phys. Lett. B}
  {\bfseries 666} (2008) 251}
  [\href{https://arxiv.org/abs/0806.3374}{{\ttfamily 0806.3374}}].

\bibitem{Liu:2012zya}
L.~Liu, K.~Orginos, F.-K.~Guo, C.~Hanhart and U.-G.~Meissner,
  \emph{{Interactions of charmed mesons with light pseudoscalar mesons from
  lattice QCD and implications on the nature of the $D_{s0}^*(2317)$}},
  \href{https://doi.org/10.1103/PhysRevD.87.014508}{\emph{Phys. Rev. D}
  {\bfseries 87} (2013) 014508}
  [\href{https://arxiv.org/abs/1208.4535}{{\ttfamily 1208.4535}}].

\bibitem{Colangelo:2003vg}
P.~Colangelo and F.~De~Fazio, \emph{{Understanding D(sJ)(2317)}},
  \href{https://doi.org/10.1016/j.physletb.2003.08.003}{\emph{Phys. Lett. B}
  {\bfseries 570} (2003) 180}
  [\href{https://arxiv.org/abs/hep-ph/0305140}{{\ttfamily hep-ph/0305140}}].

\bibitem{Dai:2003yg}
Y.-B.~Dai, C.-S.~Huang, C.~Liu and S.-L.~Zhu, \emph{{Understanding the
  D+(sJ)(2317) and D+(sJ)(2460) with sum rules in HQET}},
  \href{https://doi.org/10.1103/PhysRevD.68.114011}{\emph{Phys. Rev. D}
  {\bfseries 68} (2003) 114011}
  [\href{https://arxiv.org/abs/hep-ph/0306274}{{\ttfamily hep-ph/0306274}}].

\bibitem{Narison:2003td}
S.~Narison, \emph{{Open charm and beauty chiral multiplets in QCD}},
  \href{https://doi.org/10.1016/j.physletb.2004.11.002}{\emph{Phys. Lett. B}
  {\bfseries 605} (2005) 319}
  [\href{https://arxiv.org/abs/hep-ph/0307248}{{\ttfamily hep-ph/0307248}}].

\bibitem{Bardeen:2003kt}
W.A.~Bardeen, E.J.~Eichten and C.T.~Hill, \emph{{Chiral multiplets of heavy -
  light mesons}}, \href{https://doi.org/10.1103/PhysRevD.68.054024}{\emph{Phys.
  Rev. D} {\bfseries 68} (2003) 054024}
  [\href{https://arxiv.org/abs/hep-ph/0305049}{{\ttfamily hep-ph/0305049}}].

\bibitem{Lee:2004gt}
I.W.~Lee, T.~Lee, D.P.~Min and B.-Y.~Park, \emph{{Chiral radiative corrections
  and D(s)(2317)/D(2308) mass puzzle}},
  \href{https://doi.org/10.1140/epjc/s10052-006-0149-7}{\emph{Eur. Phys. J. C}
  {\bfseries 49} (2007) 737}
  [\href{https://arxiv.org/abs/hep-ph/0412210}{{\ttfamily hep-ph/0412210}}].

\bibitem{Wang:2006bs}
Z.G.~Wang and S.L.~Wan, \emph{{Structure of the D(s0)(2317) and the strong
  coupling constant g(D(s0)) DK with the light-cone QCD sum rules}},
  \href{https://doi.org/10.1103/PhysRevD.73.094020}{\emph{Phys. Rev. D}
  {\bfseries 73} (2006) 094020}
  [\href{https://arxiv.org/abs/hep-ph/0603007}{{\ttfamily hep-ph/0603007}}].

\bibitem{Lakhina:2006fy}
O.~Lakhina and E.S.~Swanson, \emph{{A Canonical Ds(2317)?}},
  \href{https://doi.org/10.1016/j.physletb.2007.01.075}{\emph{Phys. Lett. B}
  {\bfseries 650} (2007) 159}
  [\href{https://arxiv.org/abs/hep-ph/0608011}{{\ttfamily hep-ph/0608011}}].

\bibitem{Cheng:2003kg}
H.-Y.~Cheng and W.-S.~Hou, \emph{{B decays as spectroscope for charmed four
  quark states}},
  \href{https://doi.org/10.1016/S0370-2693(03)00834-7}{\emph{Phys. Lett. B}
  {\bfseries 566} (2003) 193}
  [\href{https://arxiv.org/abs/hep-ph/0305038}{{\ttfamily hep-ph/0305038}}].

\bibitem{Terasaki:2003qa}
K.~Terasaki, \emph{{BABAR resonance as a new window of hadron physics}},
  \href{https://doi.org/10.1103/PhysRevD.68.011501}{\emph{Phys. Rev. D}
  {\bfseries 68} (2003) 011501}
  [\href{https://arxiv.org/abs/hep-ph/0305213}{{\ttfamily hep-ph/0305213}}].

\bibitem{Chen:2004dy}
Y.-Q.~Chen and X.-Q.~Li, \emph{{A Comprehensive four-quark interpretation of
  D(s)(2317), D(s)(2457) and D(s)(2632)}},
  \href{https://doi.org/10.1103/PhysRevLett.93.232001}{\emph{Phys. Rev. Lett.}
  {\bfseries 93} (2004) 232001}
  [\href{https://arxiv.org/abs/hep-ph/0407062}{{\ttfamily hep-ph/0407062}}].

\bibitem{Maiani:2004vq}
L.~Maiani, F.~Piccinini, A.D.~Polosa and V.~Riquer, \emph{{Diquark-antidiquarks
  with hidden or open charm and the nature of X(3872)}},
  \href{https://doi.org/10.1103/PhysRevD.71.014028}{\emph{Phys. Rev. D}
  {\bfseries 71} (2005) 014028}
  [\href{https://arxiv.org/abs/hep-ph/0412098}{{\ttfamily hep-ph/0412098}}].

\bibitem{Bracco:2005kt}
M.E.~Bracco, A.~Lozea, R.D.~Matheus, F.S.~Navarra and M.~Nielsen,
  \emph{{Disentangling two- and four-quark state pictures of the charmed scalar
  mesons}}, \href{https://doi.org/10.1016/j.physletb.2005.08.037}{\emph{Phys.
  Lett. B} {\bfseries 624} (2005) 217}
  [\href{https://arxiv.org/abs/hep-ph/0503137}{{\ttfamily hep-ph/0503137}}].

\bibitem{Wang:2006uba}
Z.-G.~Wang and S.-L.~Wan, \emph{{D(s)(2317) as a tetraquark state with QCD sum
  rules in heavy quark limit}},
  \href{https://doi.org/10.1016/j.nuclphysa.2006.07.041}{\emph{Nucl. Phys. A}
  {\bfseries 778} (2006) 22}
  [\href{https://arxiv.org/abs/hep-ph/0602080}{{\ttfamily hep-ph/0602080}}].

\bibitem{Barnes:2003dj}
T.~Barnes, F.E.~Close and H.J.~Lipkin, \emph{{Implications of a DK molecule at
  2.32-GeV}}, \href{https://doi.org/10.1103/PhysRevD.68.054006}{\emph{Phys.
  Rev. D} {\bfseries 68} (2003) 054006}
  [\href{https://arxiv.org/abs/hep-ph/0305025}{{\ttfamily hep-ph/0305025}}].

\bibitem{Szczepaniak:2003vy}
A.P.~Szczepaniak, \emph{{Description of the D*(s)(2320) resonance as the D pi
  atom}}, \href{https://doi.org/10.1016/S0370-2693(03)00865-7}{\emph{Phys.
  Lett. B} {\bfseries 567} (2003) 23}
  [\href{https://arxiv.org/abs/hep-ph/0305060}{{\ttfamily hep-ph/0305060}}].

\bibitem{Kolomeitsev:2003ac}
E.E.~Kolomeitsev and M.F.M.~Lutz, \emph{{On Heavy light meson resonances and
  chiral symmetry}},
  \href{https://doi.org/10.1016/j.physletb.2003.10.118}{\emph{Phys. Lett. B}
  {\bfseries 582} (2004) 39}
  [\href{https://arxiv.org/abs/hep-ph/0307133}{{\ttfamily hep-ph/0307133}}].

\bibitem{Hofmann:2003je}
J.~Hofmann and M.F.M.~Lutz, \emph{{Open charm meson resonances with negative
  strangeness}},
  \href{https://doi.org/10.1016/j.nuclphysa.2003.12.013}{\emph{Nucl. Phys. A}
  {\bfseries 733} (2004) 142}
  [\href{https://arxiv.org/abs/hep-ph/0308263}{{\ttfamily hep-ph/0308263}}].

\bibitem{Guo:2006fu}
F.-K.~Guo, P.-N.~Shen, H.-C.~Chiang, R.-G.~Ping and B.-S.~Zou,
  \emph{{Dynamically generated 0+ heavy mesons in a heavy chiral unitary
  approach}}, \href{https://doi.org/10.1016/j.physletb.2006.08.064}{\emph{Phys.
  Lett. B} {\bfseries 641} (2006) 278}
  [\href{https://arxiv.org/abs/hep-ph/0603072}{{\ttfamily hep-ph/0603072}}].

\bibitem{Gamermann:2006nm}
D.~Gamermann, E.~Oset, D.~Strottman and M.J.~Vicente~Vacas, \emph{{Dynamically
  generated open and hidden charm meson systems}},
  \href{https://doi.org/10.1103/PhysRevD.76.074016}{\emph{Phys. Rev. D}
  {\bfseries 76} (2007) 074016}
  [\href{https://arxiv.org/abs/hep-ph/0612179}{{\ttfamily hep-ph/0612179}}].

\bibitem{Faessler:2007gv}
A.~Faessler, T.~Gutsche, V.E.~Lyubovitskij and Y.-L.~Ma, \emph{{Strong and
  radiative decays of the D(s0)*(2317) meson in the DK-molecule picture}},
  \href{https://doi.org/10.1103/PhysRevD.76.014005}{\emph{Phys. Rev. D}
  {\bfseries 76} (2007) 014005}
  [\href{https://arxiv.org/abs/0705.0254}{{\ttfamily 0705.0254}}].

\bibitem{Flynn:2007ki}
J.M.~Flynn and J.~Nieves, \emph{{Elastic s-wave B pi, D pi, D K and K pi
  scattering from lattice calculations of scalar form-factors in semileptonic
  decays}}, \href{https://doi.org/10.1103/PhysRevD.75.074024}{\emph{Phys. Rev.
  D} {\bfseries 75} (2007) 074024}
  [\href{https://arxiv.org/abs/hep-ph/0703047}{{\ttfamily hep-ph/0703047}}].

\bibitem{Guo:2009ct}
F.-K.~Guo, C.~Hanhart and U.-G.~Meissner, \emph{{Interactions between heavy
  mesons and Goldstone bosons from chiral dynamics}},
  \href{https://doi.org/10.1140/epja/i2009-10762-1}{\emph{Eur. Phys. J. A}
  {\bfseries 40} (2009) 171} [\href{https://arxiv.org/abs/0901.1597}{{\ttfamily
  0901.1597}}].

\bibitem{Guo:2015dha}
Z.-H.~Guo, U.-G.~Mei\ss{}ner and D.-L.~Yao, \emph{{New insights into the
  $D^{*}_{s0}(2317)$ and other charm scalar mesons}},
  \href{https://doi.org/10.1103/PhysRevD.92.094008}{\emph{Phys. Rev. D}
  {\bfseries 92} (2015) 094008}
  [\href{https://arxiv.org/abs/1507.03123}{{\ttfamily 1507.03123}}].

\bibitem{Albaladejo:2016hae}
M.~Albaladejo, D.~Jido, J.~Nieves and E.~Oset, \emph{{$D^*_{s0}(2317)$ and
  $\textit{DK}$ scattering in B decays from BaBar and LHCb data}},
  \href{https://doi.org/10.1140/epjc/s10052-016-4144-3}{\emph{Eur. Phys. J. C}
  {\bfseries 76} (2016) 300}
  [\href{https://arxiv.org/abs/1604.01193}{{\ttfamily 1604.01193}}].

\bibitem{Guo:2017jvc}
F.-K.~Guo, C.~Hanhart, U.-G.~Mei\ss{}ner, Q.~Wang, Q.~Zhao and B.-S.~Zou,
  \emph{{Hadronic molecules}},
  \href{https://doi.org/10.1103/RevModPhys.90.015004}{\emph{Rev. Mod. Phys.}
  {\bfseries 90} (2018) 015004}
  [\href{https://arxiv.org/abs/1705.00141}{{\ttfamily 1705.00141}}].

\bibitem{Browder:2003fk}
T.E.~Browder, S.~Pakvasa and A.A.~Petrov, \emph{{Comment on the new D(s)(*)+
  pi0 resonances}},
  \href{https://doi.org/10.1016/j.physletb.2003.10.067}{\emph{Phys. Lett. B}
  {\bfseries 578} (2004) 365}
  [\href{https://arxiv.org/abs/hep-ph/0307054}{{\ttfamily hep-ph/0307054}}].

\bibitem{vanBeveren:2003kd}
E.~van Beveren and G.~Rupp, \emph{{Observed $D_s(2317)$ and tentative
  $D(2100\text{--}2300)$ as the charmed cousins of the light scalar nonet}},
  \href{https://doi.org/10.1103/PhysRevLett.91.012003}{\emph{Phys. Rev. Lett.}
  {\bfseries 91} (2003) 012003}
  [\href{https://arxiv.org/abs/hep-ph/0305035}{{\ttfamily hep-ph/0305035}}].

\bibitem{Ortega:2016mms}
P.G.~Ortega, J.~Segovia, D.R.~Entem and F.~Fernandez, \emph{{Molecular
  components in P-wave charmed-strange mesons}},
  \href{https://doi.org/10.1103/PhysRevD.94.074037}{\emph{Phys. Rev. D}
  {\bfseries 94} (2016) 074037}
  [\href{https://arxiv.org/abs/1603.07000}{{\ttfamily 1603.07000}}].

\bibitem{Albaladejo:2018mhb}
M.~Albaladejo, P.~Fernandez-Soler, J.~Nieves and P.G.~Ortega,
  \emph{{Contribution of constituent quark model $c\bar{s}$ states to the
  dynamics of the $D_{s0}^*(2317)$ and $D_{s1}(2460)$ resonances}},
  \href{https://doi.org/10.1140/epjc/s10052-018-6176-3}{\emph{Eur. Phys. J. C}
  {\bfseries 78} (2018) 722}
  [\href{https://arxiv.org/abs/1805.07104}{{\ttfamily 1805.07104}}].

\bibitem{Bali:2003jv}
G.S.~Bali, \emph{{The D+(sJ)(2317): What can the lattice say?}},
  \href{https://doi.org/10.1103/PhysRevD.68.071501}{\emph{Phys. Rev. D}
  {\bfseries 68} (2003) 071501}
  [\href{https://arxiv.org/abs/hep-ph/0305209}{{\ttfamily hep-ph/0305209}}].

\bibitem{Dougall:2003hv}
{\scshape UKQCD} collaboration, \emph{{The Spectrum of D(s) mesons from lattice
  QCD}}, \href{https://doi.org/10.1016/j.physletb.2003.07.017}{\emph{Phys.
  Lett. B} {\bfseries 569} (2003) 41}
  [\href{https://arxiv.org/abs/hep-lat/0307001}{{\ttfamily hep-lat/0307001}}].

\bibitem{Mohler:2012na}
D.~Mohler, S.~Prelovsek and R.M.~Woloshyn, \emph{{$D \pi$ scattering and $D$
  meson resonances from lattice QCD}},
  \href{https://doi.org/10.1103/PhysRevD.87.034501}{\emph{Phys. Rev. D}
  {\bfseries 87} (2013) 034501}
  [\href{https://arxiv.org/abs/1208.4059}{{\ttfamily 1208.4059}}].

\bibitem{Mohler:2013rwa}
D.~Mohler, C.B.~Lang, L.~Leskovec, S.~Prelovsek and R.M.~Woloshyn,
  \emph{{$D_{s0}^*(2317)$ Meson and $D$-Meson-Kaon Scattering from Lattice
  QCD}}, \href{https://doi.org/10.1103/PhysRevLett.111.222001}{\emph{Phys. Rev.
  Lett.} {\bfseries 111} (2013) 222001}
  [\href{https://arxiv.org/abs/1308.3175}{{\ttfamily 1308.3175}}].

\bibitem{Lang:2014yfa}
C.B.~Lang, L.~Leskovec, D.~Mohler, S.~Prelovsek and R.M.~Woloshyn, \emph{{Ds
  mesons with DK and D*K scattering near threshold}},
  \href{https://doi.org/10.1103/PhysRevD.90.034510}{\emph{Phys. Rev. D}
  {\bfseries 90} (2014) 034510}
  [\href{https://arxiv.org/abs/1403.8103}{{\ttfamily 1403.8103}}].

\bibitem{Bali:2017pdv}
G.S.~Bali, S.~Collins, A.~Cox and A.~Sch\"afer, \emph{{Masses and decay
  constants of the $D_{s0}^*(2317)$ and $D_{s1}(2460)$ from $N_f=2$ lattice QCD
  close to the physical point}},
  \href{https://doi.org/10.1103/PhysRevD.96.074501}{\emph{Phys. Rev. D}
  {\bfseries 96} (2017) 074501}
  [\href{https://arxiv.org/abs/1706.01247}{{\ttfamily 1706.01247}}].

\bibitem{Cheung:2020mql}
{\scshape Hadron Spectrum} collaboration, \emph{{DK I = 0,$ D\overline{K} $I =
  0, 1 scattering and the $ {D}_{s0}^{\ast } $(2317) from lattice QCD}},
  \href{https://doi.org/10.1007/JHEP02(2021)100}{\emph{JHEP} {\bfseries 02}
  (2021) 100} [\href{https://arxiv.org/abs/2008.06432}{{\ttfamily
  2008.06432}}].

\bibitem{Moir:2016srx}
G.~Moir, M.~Peardon, S.M.~Ryan, C.E.~Thomas and D.J.~Wilson,
  \emph{{Coupled-Channel $D\pi$, $D\eta$ and $D_{s}\bar{K}$ Scattering from
  Lattice QCD}}, \href{https://doi.org/10.1007/JHEP10(2016)011}{\emph{JHEP}
  {\bfseries 10} (2016) 011}
  [\href{https://arxiv.org/abs/1607.07093}{{\ttfamily 1607.07093}}].

\bibitem{Gayer:2021xzv}
{\scshape Hadron Spectrum} collaboration, \emph{{Isospin-1/2 D\ensuremath{\pi}
  scattering and the lightest $ {D}_0^{\ast } $ resonance from lattice QCD}},
  \href{https://doi.org/10.1007/JHEP07(2021)123}{\emph{JHEP} {\bfseries 07}
  (2021) 123} [\href{https://arxiv.org/abs/2102.04973}{{\ttfamily
  2102.04973}}].

\bibitem{LHCb:2016lxy}
{\scshape LHCb} collaboration, \emph{{Amplitude analysis of $B^{-} \to D^{+}
  \pi^{-} \pi^{-}$ decays}},
  \href{https://doi.org/10.1103/PhysRevD.94.072001}{\emph{Phys. Rev. D}
  {\bfseries 94} (2016) 072001}
  [\href{https://arxiv.org/abs/1608.01289}{{\ttfamily 1608.01289}}].

\bibitem{LHCb:2014ioa}
{\scshape LHCb} collaboration, \emph{{Dalitz plot analysis of $B_s^0
  \rightarrow {\bar D}^0 K^- \pi^+$ decays}},
  \href{https://doi.org/10.1103/PhysRevD.90.072003}{\emph{Phys. Rev. D}
  {\bfseries 90} (2014) 072003}
  [\href{https://arxiv.org/abs/1407.7712}{{\ttfamily 1407.7712}}].

\bibitem{Liu:2023uly}
Z.-W.~Liu, J.-X.~Lu and L.-S.~Geng, \emph{{Study of the DK interaction with
  femtoscopic correlation functions}},
  \href{https://doi.org/10.1103/PhysRevD.107.074019}{\emph{Phys. Rev. D}
  {\bfseries 107} (2023) 074019}
  [\href{https://arxiv.org/abs/2302.01046}{{\ttfamily 2302.01046}}].

\bibitem{Albaladejo:2023pzq}
M.~Albaladejo, J.~Nieves and E.~Ruiz-Arriola, \emph{{Femtoscopic signatures of
  the lightest S-wave scalar open-charm mesons}},
  \href{https://doi.org/10.1103/PhysRevD.108.014020}{\emph{Phys. Rev. D}
  {\bfseries 108} (2023) 014020}
  [\href{https://arxiv.org/abs/2304.03107}{{\ttfamily 2304.03107}}].

\bibitem{Ikeno:2023ojl}
N.~Ikeno, G.~Toledo and E.~Oset, \emph{{Model independent analysis of
  femtoscopic correlation functions: An application to the
  Ds0\textasteriskcentered{}(2317)}},
  \href{https://doi.org/10.1016/j.physletb.2023.138281}{\emph{Phys. Lett. B}
  {\bfseries 847} (2023) 138281}
  [\href{https://arxiv.org/abs/2305.16431}{{\ttfamily 2305.16431}}].

\bibitem{Torres-Rincon:2023qll}
J.M.~Torres-Rincon, A.~Ramos and L.~Tolos, \emph{{Femtoscopy of D mesons and
  light mesons upon unitarized effective field theories}},
  \href{https://doi.org/10.1103/PhysRevD.108.096008}{\emph{Phys. Rev. D}
  {\bfseries 108} (2023) 096008}
  [\href{https://arxiv.org/abs/2307.02102}{{\ttfamily 2307.02102}}].

\bibitem{Molina:2009zeg}
R.~Molina, D.~Gamermann, E.~Oset and L.~Tolos, \emph{{Charm and hidden charm
  scalar mesons in the nuclear medium}},
  \href{https://doi.org/10.1140/epja/i2009-10853-y}{\emph{Eur. Phys. J. A}
  {\bfseries 42} (2009) 31} [\href{https://arxiv.org/abs/0806.3711}{{\ttfamily
  0806.3711}}].

\bibitem{Montana:2020lfi}
G.~Monta\~na, A.~Ramos, L.~Tolos and J.M.~Torres-Rincon, \emph{{Impact of a
  thermal medium on $D$ mesons and their chiral partners}},
  \href{https://doi.org/10.1016/j.physletb.2020.135464}{\emph{Phys. Lett. B}
  {\bfseries 806} (2020) 135464}
  [\href{https://arxiv.org/abs/2001.11877}{{\ttfamily 2001.11877}}].

\bibitem{Montesinos:2023qbx}
V.~Montesinos, M.~Albaladejo, J.~Nieves and L.~Tolos, \emph{{Properties of the
  $T_{cc}(3875)^+$ and $T_{\bar{c}\bar{c}}(3875)^-$ and their heavy-quark spin
  partners in nuclear matter}},
  \href{https://doi.org/10.1103/PhysRevC.108.035205}{\emph{Phys. Rev. C}
  {\bfseries 108} (2023) 035205}
  [\href{https://arxiv.org/abs/2306.17673}{{\ttfamily 2306.17673}}].

\bibitem{Bramon:2001tb}
A.~Bramon and G.~Garbarino, \emph{{Novel Bell's inequalities for entangled K0
  anti-K0 pairs}},
  \href{https://doi.org/10.1103/PhysRevLett.88.040403}{\emph{Phys. Rev. Lett.}
  {\bfseries 88} (2002) 040403}
  [\href{https://arxiv.org/abs/quant-ph/0108047}{{\ttfamily
  quant-ph/0108047}}].

\bibitem{Kaiser:1995eg}
N.~Kaiser, P.B.~Siegel and W.~Weise, \emph{{Chiral dynamics and the low-energy
  kaon - nucleon interaction}},
  \href{https://doi.org/10.1016/0375-9474(95)00362-5}{\emph{Nucl. Phys. A}
  {\bfseries 594} (1995) 325}
  [\href{https://arxiv.org/abs/nucl-th/9505043}{{\ttfamily nucl-th/9505043}}].

\bibitem{Oset:1997it}
E.~Oset and A.~Ramos, \emph{{Nonperturbative chiral approach to s wave anti-K N
  interactions}},
  \href{https://doi.org/10.1016/S0375-9474(98)00170-5}{\emph{Nucl. Phys. A}
  {\bfseries 635} (1998) 99}
  [\href{https://arxiv.org/abs/nucl-th/9711022}{{\ttfamily nucl-th/9711022}}].

\bibitem{Oller:2000fj}
J.A.~Oller and U.G.~Meissner, \emph{{Chiral dynamics in the presence of bound
  states: Kaon nucleon interactions revisited}},
  \href{https://doi.org/10.1016/S0370-2693(01)00078-8}{\emph{Phys. Lett. B}
  {\bfseries 500} (2001) 263}
  [\href{https://arxiv.org/abs/hep-ph/0011146}{{\ttfamily hep-ph/0011146}}].

\bibitem{Lutz:2001yb}
M.F.M.~Lutz and E.E.~Kolomeitsev, \emph{{Relativistic chiral SU(3) symmetry,
  large N(c) sum rules and meson baryon scattering}},
  \href{https://doi.org/10.1016/S0375-9474(01)01312-4}{\emph{Nucl. Phys. A}
  {\bfseries 700} (2002) 193}
  [\href{https://arxiv.org/abs/nucl-th/0105042}{{\ttfamily nucl-th/0105042}}].

\bibitem{Garcia-Recio:2002yxy}
C.~Garcia-Recio, J.~Nieves, E.~Ruiz~Arriola and M.J.~Vicente~Vacas, \emph{{S =
  -1 meson baryon unitarized coupled channel chiral perturbation theory and the
  S(01) Lambda(1405) and Lambda(1670) resonances}},
  \href{https://doi.org/10.1103/PhysRevD.67.076009}{\emph{Phys. Rev. D}
  {\bfseries 67} (2003) 076009}
  [\href{https://arxiv.org/abs/hep-ph/0210311}{{\ttfamily hep-ph/0210311}}].

\bibitem{Jido:2003cb}
D.~Jido, J.A.~Oller, E.~Oset, A.~Ramos and U.G.~Meissner, \emph{{Chiral
  dynamics of the two Lambda(1405) states}},
  \href{https://doi.org/10.1016/S0375-9474(03)01598-7}{\emph{Nucl. Phys. A}
  {\bfseries 725} (2003) 181}
  [\href{https://arxiv.org/abs/nucl-th/0303062}{{\ttfamily nucl-th/0303062}}].

\bibitem{Garcia-Recio:2003ejq}
C.~Garcia-Recio, M.F.M.~Lutz and J.~Nieves, \emph{{Quark mass dependence of s
  wave baryon resonances}},
  \href{https://doi.org/10.1016/j.physletb.2003.11.073}{\emph{Phys. Lett. B}
  {\bfseries 582} (2004) 49}
  [\href{https://arxiv.org/abs/nucl-th/0305100}{{\ttfamily nucl-th/0305100}}].

\bibitem{Oller:2005ig}
J.A.~Oller, J.~Prades and M.~Verbeni, \emph{{Surprises in threshold
  antikaon-nucleon physics}},
  \href{https://doi.org/10.1103/PhysRevLett.95.172502}{\emph{Phys. Rev. Lett.}
  {\bfseries 95} (2005) 172502}
  [\href{https://arxiv.org/abs/hep-ph/0508081}{{\ttfamily hep-ph/0508081}}].

\bibitem{Hyodo:2007jq}
T.~Hyodo and W.~Weise, \emph{{Effective anti-K N interaction based on chiral
  SU(3) dynamics}},
  \href{https://doi.org/10.1103/PhysRevC.77.035204}{\emph{Phys. Rev. C}
  {\bfseries 77} (2008) 035204}
  [\href{https://arxiv.org/abs/0712.1613}{{\ttfamily 0712.1613}}].

\bibitem{Hyodo:2011ur}
T.~Hyodo and D.~Jido, \emph{{The nature of the \ensuremath{\Lambda}(1405)
  resonance in chiral dynamics}},
  \href{https://doi.org/10.1016/j.ppnp.2011.07.002}{\emph{Prog. Part. Nucl.
  Phys.} {\bfseries 67} (2012) 55}
  [\href{https://arxiv.org/abs/1104.4474}{{\ttfamily 1104.4474}}].

\bibitem{Ikeda:2012au}
Y.~Ikeda, T.~Hyodo and W.~Weise, \emph{{Chiral SU(3) theory of antikaon-nucleon
  interactions with improved threshold constraints}},
  \href{https://doi.org/10.1016/j.nuclphysa.2012.01.029}{\emph{Nucl. Phys. A}
  {\bfseries 881} (2012) 98} [\href{https://arxiv.org/abs/1201.6549}{{\ttfamily
  1201.6549}}].

\bibitem{Roca:2013av}
L.~Roca and E.~Oset, \emph{{\ensuremath{\Lambda}(1405) poles obtained from
  $\pi^0 \Sigma^0$ photoproduction data}},
  \href{https://doi.org/10.1103/PhysRevC.87.055201}{\emph{Phys. Rev. C}
  {\bfseries 87} (2013) 055201}
  [\href{https://arxiv.org/abs/1301.5741}{{\ttfamily 1301.5741}}].

\bibitem{Kamiya:2016jqc}
Y.~Kamiya, K.~Miyahara, S.~Ohnishi, Y.~Ikeda, T.~Hyodo, E.~Oset et~al.,
  \emph{{Antikaon-nucleon interaction and $\Lambda$(1405) in chiral SU(3)
  dynamics}},
  \href{https://doi.org/10.1016/j.nuclphysa.2016.04.013}{\emph{Nucl. Phys. A}
  {\bfseries 954} (2016) 41}
  [\href{https://arxiv.org/abs/1602.08852}{{\ttfamily 1602.08852}}].

\bibitem{Nieves:2024dcz}
J.~Nieves, A.~Feijoo, M.~Albaladejo and M.-L.~Du, \emph{{Lowest-lying
  ${\frac{1}{2}}^-$ and ${\frac{3}{2}}^-$$\Lambda_{Q}$ resonances: from the
  strange to the bottom sectors}},
  \href{https://arxiv.org/abs/2402.12726}{{\ttfamily 2402.12726}}.

\bibitem{Akaishi:2002bg}
Y.~Akaishi and T.~Yamazaki, \emph{{Nuclear anti-K bound states in light
  nuclei}}, \href{https://doi.org/10.1103/PhysRevC.65.044005}{\emph{Phys. Rev.
  C} {\bfseries 65} (2002) 044005}.

\bibitem{Baca:2000ic}
A.~Baca, C.~Garcia-Recio and J.~Nieves, \emph{{Deeply bound levels in kaonic
  atoms}}, \href{https://doi.org/10.1016/S0375-9474(00)00152-4}{\emph{Nucl.
  Phys. A} {\bfseries 673} (2000) 335}
  [\href{https://arxiv.org/abs/nucl-th/0001060}{{\ttfamily nucl-th/0001060}}].

\bibitem{Ramos:1999ku}
A.~Ramos and E.~Oset, \emph{{The Properties of anti-K in the nuclear medium}},
  \href{https://doi.org/10.1016/S0375-9474(99)00846-5}{\emph{Nucl. Phys. A}
  {\bfseries 671} (2000) 481}
  [\href{https://arxiv.org/abs/nucl-th/9906016}{{\ttfamily nucl-th/9906016}}].

\bibitem{FINUDA:2005lqd}
{\scshape FINUDA} collaboration, \emph{{Evidence for a kaon-bound state K- p p
  produced in K- absorption reactions at rest}},
  \href{https://doi.org/10.1103/PhysRevLett.94.212303}{\emph{Phys. Rev. Lett.}
  {\bfseries 94} (2005) 212303}.

\bibitem{Oset:2005sn}
E.~Oset and H.~Toki, \emph{{A Critical analysis on deeply bound kaonic states
  in nuclei}}, \href{https://doi.org/10.1103/PhysRevC.74.015207}{\emph{Phys.
  Rev. C} {\bfseries 74} (2006) 015207}
  [\href{https://arxiv.org/abs/nucl-th/0509048}{{\ttfamily nucl-th/0509048}}].

\bibitem{Magas:2006fn}
V.K.~Magas, E.~Oset, A.~Ramos and H.~Toki, \emph{{A Critical view on the deeply
  bound K- pp system}},
  \href{https://doi.org/10.1103/PhysRevC.74.025206}{\emph{Phys. Rev. C}
  {\bfseries 74} (2006) 025206}
  [\href{https://arxiv.org/abs/nucl-th/0601013}{{\ttfamily nucl-th/0601013}}].

\bibitem{Weise:2010xn}
W.~Weise, \emph{{Antikaon interactions with nucleons and nuclei}},
  \href{https://doi.org/10.1016/j.nuclphysa.2010.01.174}{\emph{Nucl. Phys. A}
  {\bfseries 835} (2010) 51} [\href{https://arxiv.org/abs/1001.1300}{{\ttfamily
  1001.1300}}].

\bibitem{Tolos:2020aln}
L.~Tolos and L.~Fabbietti, \emph{{Strangeness in Nuclei and Neutron Stars}},
  \href{https://doi.org/10.1016/j.ppnp.2020.103770}{\emph{Prog. Part. Nucl.
  Phys.} {\bfseries 112} (2020) 103770}
  [\href{https://arxiv.org/abs/2002.09223}{{\ttfamily 2002.09223}}].

\bibitem{Tolos:2009nn}
L.~Tolos, C.~Garcia-Recio and J.~Nieves, \emph{{The Properties of D and D*
  mesons in the nuclear medium}},
  \href{https://doi.org/10.1103/PhysRevC.80.065202}{\emph{Phys. Rev. C}
  {\bfseries 80} (2009) 065202}
  [\href{https://arxiv.org/abs/0905.4859}{{\ttfamily 0905.4859}}].

\bibitem{Garcia-Recio:2010fiq}
C.~Garcia-Recio, J.~Nieves and L.~Tolos, \emph{{D mesic nuclei}},
  \href{https://doi.org/10.1016/j.physletb.2010.05.056}{\emph{Phys. Lett. B}
  {\bfseries 690} (2010) 369}
  [\href{https://arxiv.org/abs/1004.2634}{{\ttfamily 1004.2634}}].

\bibitem{Garcia-Recio:2011jcj}
C.~Garcia-Recio, J.~Nieves, L.L.~Salcedo and L.~Tolos, \emph{{$D^-$ mesic
  atoms}}, \href{https://doi.org/10.1103/PhysRevC.85.025203}{\emph{Phys. Rev.
  C} {\bfseries 85} (2012) 025203}
  [\href{https://arxiv.org/abs/1111.6535}{{\ttfamily 1111.6535}}].

\bibitem{Tolos:2006ny}
L.~Tolos, A.~Ramos and E.~Oset, \emph{{Chiral approach to antikaon s and p-wave
  interactions in dense nuclear matter}},
  \href{https://doi.org/10.1103/PhysRevC.74.015203}{\emph{Phys. Rev. C}
  {\bfseries 74} (2006) 015203}
  [\href{https://arxiv.org/abs/nucl-th/0603033}{{\ttfamily nucl-th/0603033}}].

\bibitem{Tolos:2008di}
L.~Tolos, D.~Cabrera and A.~Ramos, \emph{{Strange mesons in nuclear matter at
  finite temperature}},
  \href{https://doi.org/10.1103/PhysRevC.78.045205}{\emph{Phys. Rev. C}
  {\bfseries 78} (2008) 045205}
  [\href{https://arxiv.org/abs/0807.2947}{{\ttfamily 0807.2947}}].

\bibitem{Cabrera:2014lca}
D.~Cabrera, L.~Tol\'os, J.~Aichelin and E.~Bratkovskaya, \emph{{Antistrange
  meson-baryon interaction in hot and dense nuclear matter}},
  \href{https://doi.org/10.1103/PhysRevC.90.055207}{\emph{Phys. Rev. C}
  {\bfseries 90} (2014) 055207}
  [\href{https://arxiv.org/abs/1406.2570}{{\ttfamily 1406.2570}}].

\bibitem{Albaladejo:2021cxj}
M.~Albaladejo, J.M.~Nieves and L.~Tolos, \emph{{$D\overline{D}{}^\ast$
  scattering and $\chi_{c1}(3872)$ in nuclear matter}},
  \href{https://doi.org/10.1103/PhysRevC.104.035203}{\emph{Phys. Rev. C}
  {\bfseries 104} (2021) 035203}
  [\href{https://arxiv.org/abs/2102.08589}{{\ttfamily 2102.08589}}].

\bibitem{Tolos:2013gta}
L.~Tolos, \emph{{Charming mesons with baryons and nuclei}},
  \href{https://doi.org/10.1142/S0218301313300270}{\emph{Int. J. Mod. Phys. E}
  {\bfseries 22} (2013) 1330027}
  [\href{https://arxiv.org/abs/1309.7305}{{\ttfamily 1309.7305}}].

\bibitem{Garcia-Recio:2008rjt}
C.~Garcia-Recio, V.K.~Magas, T.~Mizutani, J.~Nieves, A.~Ramos, L.L.~Salcedo
  et~al., \emph{{The s-wave charmed baryon resonances from a coupled-channel
  approach with heavy quark symmetry}},
  \href{https://doi.org/10.1103/PhysRevD.79.054004}{\emph{Phys. Rev. D}
  {\bfseries 79} (2009) 054004}
  [\href{https://arxiv.org/abs/0807.2969}{{\ttfamily 0807.2969}}].

\bibitem{Gamermann:2010zz}
D.~Gamermann, C.~Garcia-Recio, J.~Nieves, L.L.~Salcedo and L.~Tolos,
  \emph{{Exotic dynamically generated baryons with negative charm quantum
  number}}, \href{https://doi.org/10.1103/PhysRevD.81.094016}{\emph{Phys. Rev.
  D} {\bfseries 81} (2010) 094016}
  [\href{https://arxiv.org/abs/1002.2763}{{\ttfamily 1002.2763}}].

\bibitem{Romanets:2012hm}
O.~Romanets, L.~Tolos, C.~Garcia-Recio, J.~Nieves, L.L.~Salcedo and
  R.G.E.~Timmermans, \emph{{Charmed and strange baryon resonances with
  heavy-quark spin symmetry}},
  \href{https://doi.org/10.1103/PhysRevD.85.114032}{\emph{Phys. Rev. D}
  {\bfseries 85} (2012) 114032}
  [\href{https://arxiv.org/abs/1202.2239}{{\ttfamily 1202.2239}}].

\bibitem{Nieves:2017lij}
J.~Nieves and J.E.~Sobczyk, \emph{{In medium dispersion relation effects in
  nuclear inclusive reactions at intermediate and low energies}},
  \href{https://doi.org/10.1016/j.aop.2017.06.002}{\emph{Annals Phys.}
  {\bfseries 383} (2017) 455}
  [\href{https://arxiv.org/abs/1701.03628}{{\ttfamily 1701.03628}}].

\bibitem{Aceti:2014ala}
F.~Aceti, L.R.~Dai, L.S.~Geng, E.~Oset and Y.~Zhang, \emph{{Meson-baryon
  components in the states of the baryon decuplet}},
  \href{https://doi.org/10.1140/epja/i2014-14057-2}{\emph{Eur. Phys. J. A}
  {\bfseries 50} (2014) 57} [\href{https://arxiv.org/abs/1301.2554}{{\ttfamily
  1301.2554}}].

\bibitem{Hyodo:2013nka}
T.~Hyodo, \emph{{Structure and compositeness of hadron resonances}},
  \href{https://doi.org/10.1142/S0217751X13300457}{\emph{Int. J. Mod. Phys. A}
  {\bfseries 28} (2013) 1330045}
  [\href{https://arxiv.org/abs/1310.1176}{{\ttfamily 1310.1176}}].

\bibitem{Sekihara:2014kya}
T.~Sekihara, T.~Hyodo and D.~Jido, \emph{{Comprehensive analysis of the wave
  function of a hadronic resonance and its compositeness}},
  \href{https://doi.org/10.1093/ptep/ptv081}{\emph{PTEP} {\bfseries 2015}
  (2015) 063D04} [\href{https://arxiv.org/abs/1411.2308}{{\ttfamily
  1411.2308}}].

\bibitem{Albaladejo:2015kea}
M.~Albaladejo, M.~Nielsen and E.~Oset, \emph{{$D_{s0}^{\ast\pm}(2317)$ and $KD$
  scattering from $B^0_s$ decay}},
  \href{https://doi.org/10.1016/j.physletb.2015.05.019}{\emph{Phys. Lett. B}
  {\bfseries 746} (2015) 305}
  [\href{https://arxiv.org/abs/1501.03455}{{\ttfamily 1501.03455}}].

\bibitem{Nieves:1999bx}
J.~Nieves and E.~Ruiz~Arriola, \emph{{Bethe-Salpeter approach for unitarized
  chiral perturbation theory}},
  \href{https://doi.org/10.1016/S0375-9474(00)00321-3}{\emph{Nucl. Phys. A}
  {\bfseries 679} (2000) 57}
  [\href{https://arxiv.org/abs/hep-ph/9907469}{{\ttfamily hep-ph/9907469}}].

\bibitem{Gamermann:2009uq}
D.~Gamermann, J.~Nieves, E.~Oset and E.~Ruiz~Arriola, \emph{{Couplings in
  coupled channels versus wave functions: application to the X(3872)
  resonance}}, \href{https://doi.org/10.1103/PhysRevD.81.014029}{\emph{Phys.
  Rev. D} {\bfseries 81} (2010) 014029}
  [\href{https://arxiv.org/abs/0911.4407}{{\ttfamily 0911.4407}}].

\bibitem{Matuschek:2020gqe}
I.~Matuschek, V.~Baru, F.-K.~Guo and C.~Hanhart, \emph{{On the nature of
  near-threshold bound and virtual states}},
  \href{https://doi.org/10.1140/epja/s10050-021-00413-y}{\emph{Eur. Phys. J. A}
  {\bfseries 57} (2021) 101}
  [\href{https://arxiv.org/abs/2007.05329}{{\ttfamily 2007.05329}}].

\bibitem{Esposito:2021vhu}
A.~Esposito, L.~Maiani, A.~Pilloni, A.D.~Polosa and V.~Riquer, \emph{{From the
  line shape of the X(3872) to its structure}},
  \href{https://doi.org/10.1103/PhysRevD.105.L031503}{\emph{Phys. Rev. D}
  {\bfseries 105} (2022) L031503}
  [\href{https://arxiv.org/abs/2108.11413}{{\ttfamily 2108.11413}}].

\bibitem{Li:2021cue}
Y.~Li, F.-K.~Guo, J.-Y.~Pang and J.-J.~Wu, \emph{{Generalization of
  Weinberg\textquoteright{}s compositeness relations}},
  \href{https://doi.org/10.1103/PhysRevD.105.L071502}{\emph{Phys. Rev. D}
  {\bfseries 105} (2022) L071502}
  [\href{https://arxiv.org/abs/2110.02766}{{\ttfamily 2110.02766}}].

\bibitem{Song:2022yvz}
J.~Song, L.R.~Dai and E.~Oset, \emph{{How much is the compositeness of a bound
  state constrained by a and $r_0$? The role of the interaction range}},
  \href{https://doi.org/10.1140/epja/s10050-022-00753-3}{\emph{Eur. Phys. J. A}
  {\bfseries 58} (2022) 133}
  [\href{https://arxiv.org/abs/2201.04414}{{\ttfamily 2201.04414}}].

\bibitem{Baru:2003qq}
V.~Baru, J.~Haidenbauer, C.~Hanhart, Y.~Kalashnikova and A.E.~Kudryavtsev,
  \emph{{Evidence that the a(0)(980) and f(0)(980) are not elementary
  particles}},
  \href{https://doi.org/10.1016/j.physletb.2004.01.088}{\emph{Phys. Lett. B}
  {\bfseries 586} (2004) 53}
  [\href{https://arxiv.org/abs/hep-ph/0308129}{{\ttfamily hep-ph/0308129}}].

\bibitem{Baru:2010ww}
V.~Baru, C.~Hanhart, Y.S.~Kalashnikova, A.E.~Kudryavtsev and A.V.~Nefediev,
  \emph{{Interplay of quark and meson degrees of freedom in a near-threshold
  resonance}}, \href{https://doi.org/10.1140/epja/i2010-10929-7}{\emph{Eur.
  Phys. J. A} {\bfseries 44} (2010) 93}
  [\href{https://arxiv.org/abs/1001.0369}{{\ttfamily 1001.0369}}].

\bibitem{Hanhart:2011jz}
C.~Hanhart, Y.S.~Kalashnikova and A.V.~Nefediev, \emph{{Interplay of quark and
  meson degrees of freedom in a near-threshold resonance: multi-channel case}},
  \href{https://doi.org/10.1140/epja/i2011-11101-9}{\emph{Eur. Phys. J. A}
  {\bfseries 47} (2011) 101} [\href{https://arxiv.org/abs/1106.1185}{{\ttfamily
  1106.1185}}].

\bibitem{Aceti:2012dd}
F.~Aceti and E.~Oset, \emph{{Wave functions of composite hadron states and
  relationship to couplings of scattering amplitudes for general partial
  waves}}, \href{https://doi.org/10.1103/PhysRevD.86.014012}{\emph{Phys. Rev.
  D} {\bfseries 86} (2012) 014012}
  [\href{https://arxiv.org/abs/1202.4607}{{\ttfamily 1202.4607}}].

\bibitem{Hyodo:2011qc}
T.~Hyodo, D.~Jido and A.~Hosaka, \emph{{Compositeness of dynamically generated
  states in a chiral unitary approach}},
  \href{https://doi.org/10.1103/PhysRevC.85.015201}{\emph{Phys. Rev. C}
  {\bfseries 85} (2012) 015201}
  [\href{https://arxiv.org/abs/1108.5524}{{\ttfamily 1108.5524}}].

\bibitem{Kamiya:2015aea}
Y.~Kamiya and T.~Hyodo, \emph{{Structure of near-threshold quasibound states}},
  \href{https://doi.org/10.1103/PhysRevC.93.035203}{\emph{Phys. Rev. C}
  {\bfseries 93} (2016) 035203}
  [\href{https://arxiv.org/abs/1509.00146}{{\ttfamily 1509.00146}}].

\bibitem{Garcia-Recio:2015jsa}
C.~Garcia-Recio, C.~Hidalgo-Duque, J.~Nieves, L.L.~Salcedo and L.~Tolos,
  \emph{{Compositeness of the strange, charm, and beauty odd parity $\Lambda$
  states}}, \href{https://doi.org/10.1103/PhysRevD.92.034011}{\emph{Phys. Rev.
  D} {\bfseries 92} (2015) 034011}
  [\href{https://arxiv.org/abs/1506.04235}{{\ttfamily 1506.04235}}].

\bibitem{Guo:2015daa}
Z.-H.~Guo and J.A.~Oller, \emph{{Probabilistic interpretation of compositeness
  relation for resonances}},
  \href{https://doi.org/10.1103/PhysRevD.93.096001}{\emph{Phys. Rev. D}
  {\bfseries 93} (2016) 096001}
  [\href{https://arxiv.org/abs/1508.06400}{{\ttfamily 1508.06400}}].

\bibitem{Kamiya:2016oao}
Y.~Kamiya and T.~Hyodo, \emph{{Generalized weak-binding relations of
  compositeness in effective field theory}},
  \href{https://doi.org/10.1093/ptep/ptw188}{\emph{PTEP} {\bfseries 2017}
  (2017) 023D02} [\href{https://arxiv.org/abs/1607.01899}{{\ttfamily
  1607.01899}}].

\bibitem{Sekihara:2016xnq}
T.~Sekihara, \emph{{Two-body wave functions and compositeness from scattering
  amplitudes. I. General properties with schematic models}},
  \href{https://doi.org/10.1103/PhysRevC.95.025206}{\emph{Phys. Rev. C}
  {\bfseries 95} (2017) 025206}
  [\href{https://arxiv.org/abs/1609.09496}{{\ttfamily 1609.09496}}].

\bibitem{Oller:2017alp}
J.A.~Oller, \emph{{New results from a number operator interpretation of the
  compositeness of bound and resonant states}},
  \href{https://doi.org/10.1016/j.aop.2018.07.023}{\emph{Annals Phys.}
  {\bfseries 396} (2018) 429}
  [\href{https://arxiv.org/abs/1710.00991}{{\ttfamily 1710.00991}}].

\bibitem{Kinugawa:2021ykv}
T.~Kinugawa and T.~Hyodo, \emph{{Role of the effective range in the
  weak-binding relation}},
  \href{https://doi.org/10.1051/epjconf/202226201019}{\emph{EPJ Web Conf.}
  {\bfseries 262} (2022) 01019}
  [\href{https://arxiv.org/abs/2112.00249}{{\ttfamily 2112.00249}}].

\bibitem{Sazdjian:2022kaf}
H.~Sazdjian, \emph{{The Interplay between Compact and Molecular Structures in
  Tetraquarks}}, \href{https://doi.org/10.3390/sym14030515}{\emph{Symmetry}
  {\bfseries 14} (2022) 515}
  [\href{https://arxiv.org/abs/2202.01081}{{\ttfamily 2202.01081}}].

\bibitem{Song:2023pdq}
J.~Song, L.R.~Dai and E.~Oset, \emph{{Evolution of compact states to molecular
  ones with coupled channels: The case of the X(3872)}},
  \href{https://doi.org/10.1103/PhysRevD.108.114017}{\emph{Phys. Rev. D}
  {\bfseries 108} (2023) 114017}
  [\href{https://arxiv.org/abs/2307.02382}{{\ttfamily 2307.02382}}].

\bibitem{Dai:2023kwv}
L.R.~Dai, J.~Song and E.~Oset, \emph{{Evolution of genuine states to molecular
  ones: The Tcc(3875) case}},
  \href{https://doi.org/10.1016/j.physletb.2023.138200}{\emph{Phys. Lett. B}
  {\bfseries 846} (2023) 138200}
  [\href{https://arxiv.org/abs/2306.01607}{{\ttfamily 2306.01607}}].

\bibitem{BaBar:2014jjr}
{\scshape BaBar} collaboration, \emph{{Dalitz plot analyses of $B^0 \to D^0D^-
  K^+$ and $B^+ \to \overline{D}^0D^0K^+$ decays}},
  \href{https://doi.org/10.1103/PhysRevD.91.052002}{\emph{Phys. Rev. D}
  {\bfseries 91} (2015) 052002}
  [\href{https://arxiv.org/abs/1412.6751}{{\ttfamily 1412.6751}}].

\bibitem{Garcia-Recio:2010enl}
C.~Garcia-Recio, L.S.~Geng, J.~Nieves and L.L.~Salcedo, \emph{{Low-lying even
  parity meson resonances and spin-flavor symmetry}},
  \href{https://doi.org/10.1103/PhysRevD.83.016007}{\emph{Phys. Rev. D}
  {\bfseries 83} (2011) 016007}
  [\href{https://arxiv.org/abs/1005.0956}{{\ttfamily 1005.0956}}].

\bibitem{Hirenzaki:2000da}
S.~Hirenzaki, Y.~Okumura, H.~Toki, E.~Oset and A.~Ramos, \emph{{Chiral unitary
  model for the kaonic atom}},
  \href{https://doi.org/10.1103/PhysRevC.61.055205}{\emph{Phys. Rev. C}
  {\bfseries 61} (2000) 055205}.

\end{thebibliography}\endgroup

\end{document}